\documentclass[lettersize,journal]{IEEEtran}

\usepackage{amsmath,amssymb,amsfonts}
\usepackage{algorithmic}
\usepackage{algorithm}
\usepackage{graphicx}
\usepackage{textcomp}
\usepackage{epstopdf}
\usepackage{caption}
\usepackage{array}
\usepackage{hyperref}
\usepackage{stfloats}
\usepackage{setspace}
\usepackage{float}
\usepackage{balance}
\usepackage{glossaries}
\usepackage{subcaption}
\usepackage{booktabs}
\usepackage{diagbox}
\usepackage{multirow}
\usepackage{makecell}
\usepackage{longtable}
\usepackage{siunitx}

\usepackage[justification=centering]{caption}
 
\usepackage{url}
\usepackage[table]{xcolor}
\definecolor{Red}{rgb}{0.9,0,0}

\newtheorem{lemma}{\textbf{Lemma}}

\newtheorem{theorem}{\textbf{Theorem}}

\newtheorem{remark}{\textbf{Remark}}
\allowdisplaybreaks
\ifCLASSINFOpdf

\else

\fi

\begin{document}
	
\title{ Dynamic  Modeling  and  Control for an Offshore Semisubmersible Floating Wind Turbine}

	\author{Yingjie Gong,
		Qinmin Yang, \IEEEmembership{Senior Member, IEEE}, Hua Geng, \IEEEmembership{Fellow, IEEE}, Wenchao Meng, \IEEEmembership{Senior Member, IEEE}, Lin Wang
		\thanks{Y. Gong and W. Meng are  with the State Key Laboratory of Industrial Control Technology, College of Control Science and Engineering, Zhejiang University, Hangzhou 310027, China.}
		\thanks{Q. Yang is with  the State Key Laboratory of Industrial Control Technology, College of Control Science and Engineering, and the Huzhou Institute, Zhejiang University, Hangzhou 310027, China.}
\thanks{H. Geng is with the Department of Automation, Beijing National Research Center for Information Science and Technology, Tsinghua University, Beijing 100084, China.}
\thanks{L. Wang is with the  Zhejiang Windey Co. Ltd., Hangzhou 310027, China.}
 }
	
	\maketitle

	\IEEEpeerreviewmaketitle

	\begin{abstract}
Floating wind turbines (FWTs) hold significant potential for the exploitation of offshore renewable energy resources.
Nevertheless, prior to the construction of FWTs, it is imperative to tackle several critical challenges, especially the issue of performance degradation under combined wind and wave loads.
This study initiates with the development of a simplified nonlinear dynamical model for a semi-submersible FWT.
In particular,  both the rotor dynamics and the finite rotations of the platform are considered in presented modeling approach, thereby effectively capturing the complex interplay between the platform, tower, nacelle, and rotor under combined wind and wave loads.
Subsequently, based on the developed FWT  model, a novel adaptive nonlinear pitch controller is formulated with the goal of striking a trade-off between regulating power generation and reducing platform motion.
 Notably, the proposed control strategy adopts a continuous control approach, strategically beneficial in circumventing the chattering phenomenon commonly associated with sliding mode control. Furthermore, the controller integrates an online approximator and a robust integral of the sign of the tracking error, facilitating real-time learning of system unknown dynamics while compensating for bounded disturbances.
 Finally,  both the accuracy of the established nonlinear FWT model in predicting key dynamics and the superiority of the presented pitch controller are validated through comprehensive comparative studies.
	\end{abstract}
	
\def\abstractname{Note to Practitioners}
\begin{abstract}
	This paper addresses the conflicting goals between power regulation and load mitigation for floating wind turbines (FWTs) to ensure the reliable operation of wind turbine systems.
	This remains an ongoing challenge due to the inherent complexity of existing FWT models,
	frequently resulting in controllers  crafted using linearized representations that fail to accommodate real-world uncertainties effectively.
	Through the utilization of a simplified physical-based nonlinear FWT model, a novel adaptive nonlinear pitch controller emerges as a promising solution.
	Notably, the developed nonlinear FWT model elucidates the coupling between rotor and platform degrees of freedom clearly and succinctly, facilitating the design of intelligent controllers.
	Our approach demonstrates the capability to concurrently regulate power production and stabilize the platform.
	Additionally, an online approximator is integrated into the controller to capture system dynamics, thus augmenting adaptability and diminishing reliance on high-gain feedback compensation. Importantly, this control strategy holds promise for extension and implementation in various other renewable energy systems.
\end{abstract}


	\section{Introduction}

\IEEEPARstart{G}{iven}
the predominant concentration of wind energy in deep-sea areas, the last few decades have witnessed a swift rise in the advancement of floating wind turbines (FWTs) \cite{musial2019,xie2023power,song2023fault}. Unlike bottom-fixed  WTs, FWTs experience a more intricate dynamic behavior due to the combined influence of wind and wave loads. Moreover, the dynamic performance of  FWTs may undergo notable degradation due to the motion of the floating platform, even when mooring lines anchor the FWT to the seabed \cite{ren2021offshore}.
Therefore, devising adaptive and robust control strategies that improve power output while minimizing platform motion remains a primary challenge in FWTs.

Initially, the characteristic analysis of FWTs under wind and wave loads   is pivotal in devising a well-designed controller,
which highlights that  an effective mathematical model is the foundation  for illustrating the nonlinear dynamics of FWTs in controller design.
Various studies have explored modeling approaches for different types of FWTs, such as the high-fidelity fatigue, aerodynamics, structures, and turbulence (FAST) model proposed by the National Renewable Energy Laboratory (NREL) \cite{jonkman2005fast}. This model excels in detailing  FWTs' dynamics, but its complexity renders it impractical for controller design applications.
By contrast, simplified models, striking a judicious balance between conciseness and accuracy, have emerged as a more pragmatic option.
In \cite{stewart2012determining, stewart2013offshore}, the 2-degree-of-freedom (2-DOF) reduced-order models were introduced for barge, TLP, and spar types of FWTs, while a 3-DOF dynamic model was developed in \cite{betti2013development} specifically for a barge-type FWT.
 Then to address out-of-plane  motions and reduce fatigue damage, a control-oriented rigid offshore WT was introduced in \cite{homer2017physics} based on the first principles, and subsequent consideration of structural flexibility was presented in \cite{song2023control}.

 Nonetheless, it is important to note that the prevalent modeling approaches predominantly presume a static or constant rotor rotational speed, overlooking the fact that dynamic variations in rotor speed significantly affect the aerodynamic loads and the rotor's gyroscopic moment, thereby altering the system's overall responses. Consequently, to improve the accuracy of the FWT system's dynamic behaviors, integrating rotor dynamics into the system's overall dynamics is essential \cite{al2019dynamic}.
Furthermore, despite expectations of increased focus on semi-submersible FWTs (SSFWTs) in future developments \cite{musial2019}, research on the coupled dynamic models of SSFWTs remains limited, especially when compared to other FWT designs,  particularly the spar-type FWTs. Consequently, there is a critical need to develop a comprehensive dynamic model for SSFWTs that incorporates rotor dynamics. This model will facilitate more effective control design strategies, ultimately improving controller performance.

In addition, the main control goal for FWTs in region III is to regulate power production at its rated value by adjusting  blade pitch angles \cite{li2024rotor}.
However,  the direct application of traditional controller originally designed for
``standard onshore" WTs tends to be inefficient for FWTs, which is primarily attributed to the additive motions of floating platforms \cite{larsen2007method,lee2019effects}.
Thus, it is important for FWTs to enhance the system damping and stabilize overall performance.
In \cite{jonkman2008influence}, extensive analyses of blade pitch control had been conducted on three main platforms, which effectively mitigated power fluctuations and platform motion.
Meanwhile, some advanced control algorithms, such as linear quadratic regulator-based controllers \cite{namik2010individual,sarkar2020individual} and model predictive control \cite{evans2014robust} have been  examined to address the damping of  FWTs.
Although these methods can achieve better power regulation and structural load mitigation than the baseline  controller \cite{bossanyi2003gh},
an obvious limitation of these methods is the reliance on the conventional linear models as a prevalent routine, and thus
exhibit performance degradation   outside the linearization points \cite{sarkar2018development,tang2022wind}.

An effective solution to address these limitations is  to  use  nonlinear control theory.
Notably, a nonlinear controller known as sliding mode control (SMC) has recently been implemented in the context of vibration mitigation for a barge-type FWT in \cite{shah2021advanced}.
It is crucial to highlight that SMC stands out as one of the most potent nonlinear control techniques documented in the literature, adept at addressing uncertainties and disturbances stemming from wind-wave joint excitation.
However, despite its commendable robustness, SMC is plagued by the well-documented chattering phenomenon resulting from its discontinuous behavior.
 Furthermore, the sliding mode controller typically employs conservative control gains to ensure effective performance in the presence of disturbances.
To augment control performance, a second-order SMC based control algorithm, reducing the chattering,  has been applied in \cite{zhang2019adaptive,zhang2021individual,gutierrez2022simplified} to
achieve  power regulation for spar-type FWTs.
Unfortunately, a significant drawback
lies in the reliance on applied the linear model framework in this approach, treating the FWT system as a black box for direct execution of SMC. This oversight neglects detailed wind/wave descriptions and known system parts, resulting in excessive SMC effects that amplify damage to the pitch actuator.

Motivated by the aforementioned issues, the primary objective of this paper is to formulate a  simplified nonlinear model for SSFWTs. Subsequently, leveraging the established model, we design a robust adaptive pitch controller employing a continuous control scheme.
The contributions of this paper can be summarized as:
\begin{enumerate}
    \item An improved modeling approach for SSFWT is proposed  considering rotor dynamics and finite platform rotations.
     The resulting model is succinctly presented in matrix form with clearly defined elements, facilitating more effective control strategies design and ultimately enhancing controller performance.
    \item  Based on the established model, a new robust nonlinear pitch controller  has been developed  to achieve a trade-off between the rotor speed regulation and suppressing the platform motion of SSFWT.
	\item  In comparison to exist nonlinear controllers, such as SMC, an online approximator  is utilized to learn the  unknown system dynamics, thereby enhancing the controller's adaptability and reducing the reliance on high-gain feedback compensation.
\end{enumerate}

This paper is structured as follows: Section II introduces the modeling of SSFWT. In Section III, we present a novel blade pitch controller and then demonstrate its stability. Section IV shows verification results conducted on a SSFWT, followed by the conclusion of this  study is given in Section V.


%
%


	\section{Modeling of the SSFWT system }
\begin{figure}[!htb]
	\begin{center}
		\centerline{\includegraphics[width=0.450\textwidth]{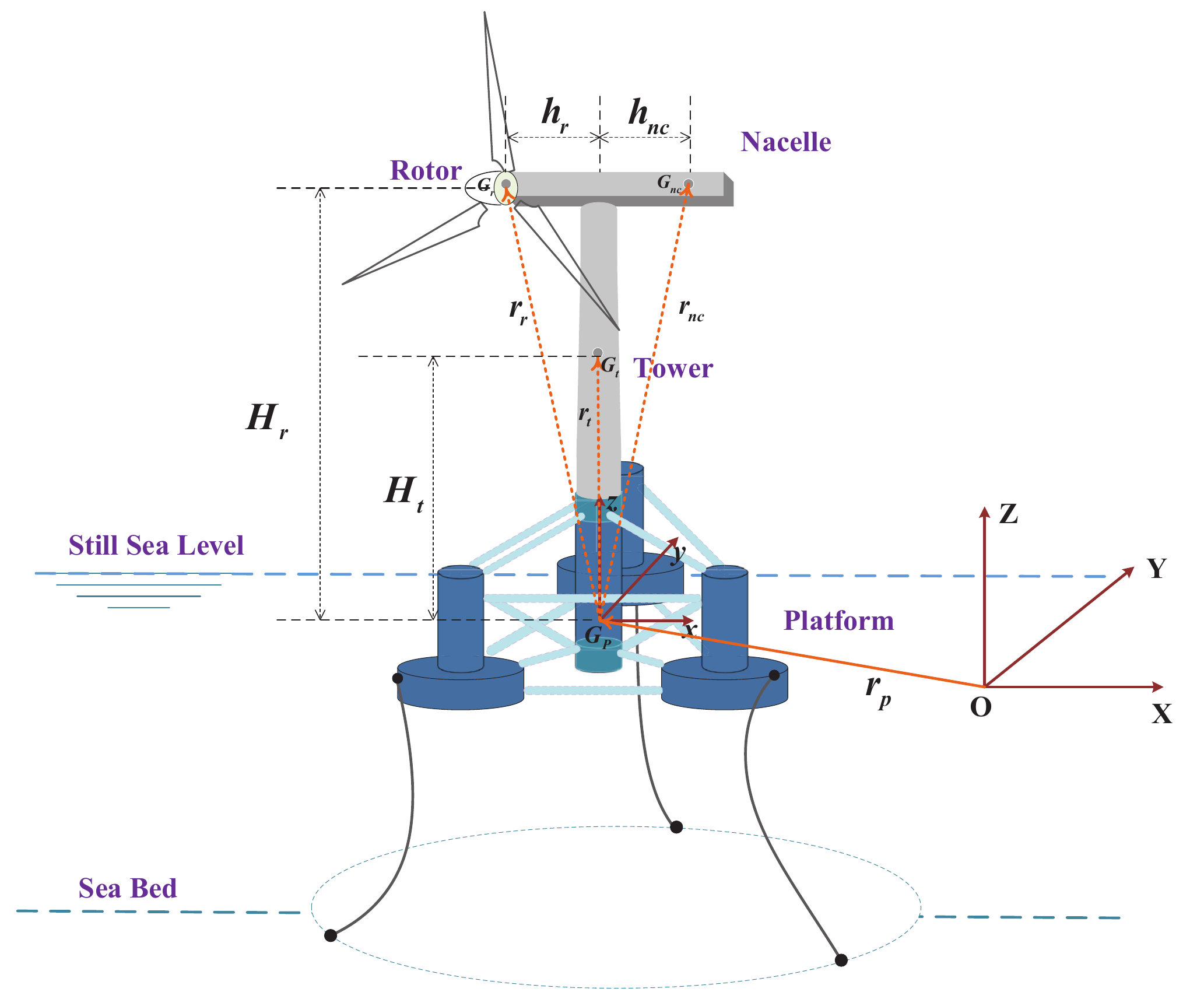}}
		\caption{Schematic diagram of the SSFWT. }
		\label{fig0}
	\end{center}
\end{figure}

The schematic diagram of the SSFWT  is depicted in Fig. \ref{fig0}. Meanwhile,
the dynamic characteristics of  SSFWT are intricately influenced by four principal mechanical components:\
\begin{enumerate}
		\item \textbf{Floating Platform}:\ Characterized by mass $m_{p}$, center of mass $G_{p}$   and inertia tensor  $\mathbf{I}_{\mathrm{p}}=\operatorname{diag}\{I_{p_x}, I_{ p_y}, I_{p_z}\}$.
		\item  \textbf{Tower}: \  Comprising mass $m_t$, center of mass $G_{t}$   and inertia tensor $\mathbf{I}_{\mathrm{t}}=\operatorname{diag}\{I_{t_x}, I_{t_y}, I_{t_z}\}$. The tower supports the rotor-nacelle assembly, and the yaw motion is not considered.
		\item \textbf{Nacelle}: \ Featuring mass $m_{n_c}$, center of mass $G_{n_c}$  and inertia tensor $\mathbf{I}_{\mathrm{n}}=\operatorname{diag}\{I_{n_x}, I_{n_y}, I_{n_z}\}$. 	
		\item   \textbf{Rotor}: \   Represented as a rigid disc with angular velocity  $\Omega_r(t)$,  mass  $m_r$,  center of mass $G_{r}$  and inertia tensor $\mathbf{I}_{\mathrm{r}}=\operatorname{diag}\{I_{r_x}, I_{r_y}, I_{r_z}\}$.
	\end{enumerate}
Note that two right-handed Cartesian coordinate frames, $\mathcal{F}_I$ (inertial frame) and $\mathcal{F}_p$ (body-fixed frame), are defined in Fig. \ref{fig0}.
 The origin of $\mathcal{F}_I$ is located at the platform's center of gravity ($\mathrm{G}_p$) before any displacement, with its unit vector $[\mathbf{\hat{e}}_3]_I$ oriented vertically upward. The unit vectors $[\mathbf{\hat{e}}_1]_I$ and $[\mathbf{\hat{e}}_2]_I$ are horizontally aligned, parallel to the Earth's surface. Conversely, $\mathcal{F}_p$ is affixed to $\mathrm{G}_p$ and initially coincides with $\mathcal{F}_I$ in the absence of wind or wave disturbances. It's noteworthy that the subscript denotes the vector representation frame in this analysis, whereas vectors lacking superscripts denote representation in any arbitrary frame. Furthermore, the principal axes associated with the inertia tensor of rigid components align with $\mathcal{F}_p$.

 \subsection{System kinematics}
The floating platform   is typically exhibits six DOFs in frame $\mathcal{F}_I$: three for translation (surge, sway, and heave) and three for rotation (roll, pitch, and yaw), denoted by
 \[
 	\mathbf{r}_p=\left[\begin{array}{lll}r_x & r_y & r_z\end{array}\right]^T, \quad
 \boldsymbol{\theta}_p=\left[\begin{array}{lll}\theta_{{x}} &\theta_{{y}} & \theta_{{z}} \end{array}\right]^T
 \]
 where   $\mathbf{r}_p$ represents translational displacements of the origin of frame $\mathcal{F}_p$ with respect to  frame $\mathcal{F}_I$, expressed in $\mathcal{F}_I$. Meanwhile, the rotational angles are characterized by the Euler angle triad $\boldsymbol{\theta}_p$.
The 3-2-1 Euler angle sequence \cite{junkins2012optimal} characterizes finite rotation in this paper. The rotation matrix $\mathbf{R}$ in \eqref{eq-1} facilitates the transformation of vectors between frames $\mathcal{F}_p$ and $\mathcal{F}_I$.
\begin{equation}\label{eq-1}
	\mathbf{R}\hspace{-0.5mm}=\hspace{-1.0mm}\left[\begin{array}{lll}
	\hspace{-1.5mm}	c_{\theta_{{\mathrm{z}}}} c_{\theta_{{\mathrm{y}}}} \hspace{-1.0mm}&\hspace{-1.0mm} c_{\theta_{{\mathrm{z}}}} s_{\theta_{{\mathrm{y}}}} s_{\theta_{{\mathrm{x}}}}-s_{\theta_{{\mathrm{z}}}} c_{\theta_{{\mathrm{x}}}} \hspace{-1.0mm}&\hspace{-1.0mm} c_{\theta_{{\mathrm{z}}}} s_{\theta_{{\mathrm{y}}}} c_{\theta_{{\mathrm{x}}}}+s_{\theta_{{\mathrm{x}}}} s_{\theta_{{\mathrm{z}}}}\hspace{-1.5mm} \\
	\hspace{-1.5mm}	s_{\theta_{{\mathrm{z}}}} c_{\theta_{{\mathrm{y}}}} \hspace{-1.0mm}&\hspace{-1.0mm} s_{\theta_{{\mathrm{z}}}} s_{ \theta_{{\mathrm{y}}}} s_{\theta_{{\mathrm{x}}}}+c_{\theta_{{\mathrm{z}}}} c_{\theta_{{\mathrm{x}}}} \hspace{-1.0mm}&\hspace{-1.0mm} s_{\theta_{{\mathrm{z}}}}  s_{\theta_{{\mathrm{y}}}} c_{\theta_{{\mathrm{x}}}}-c_{\theta_{{\mathrm{z}}}} s_{\theta_{{\mathrm{x}}}} \hspace{-1.5mm}\\
	\hspace{-1.5mm}	-s_{\theta_{{\mathrm{y}}}} \hspace{-1.0mm}&\hspace{-1.0mm} c_{\theta_{{\mathrm{y}}}} s_{\theta_{{\mathrm{x}}}} \hspace{-1.0mm}&\hspace{-1.0mm} c_{\theta_{{\mathrm{y}}}} c_{ \theta_{{\mathrm{x}}}}\hspace{-1.5mm}
	\end{array}\right]
\end{equation}
where $c_{(\cdot)}=\cos(\cdot)$ and  $z_{(\cdot)}=\sin(\cdot)$.
Subsequently, the translational velocity of the platform within the frame $ \mathcal{F}_p$
 can be denoted as
\begin{equation}\label{eq-2}
 [\mathbf{v}_\mathrm{p}]_p=\left[v_{x}, v_{y}, v_{z} \right]^{\mathrm{T}}=\mathbf{\mathbf{R}}^{\mathrm{T}}   [\mathbf{\dot{r}}_p]_I
\end{equation}
Also, the Euler kinematical equation of the 3-2-1
Euler angle sequence is
\begin{equation}\label{eq-3}
	[\mathbf{\dot{\boldsymbol{\theta}}}_p]_I =\mathbf{J} [\boldsymbol{\omega}_\mathrm{p}]_p  =\left[\begin{array}{lll}
	 	1 & \frac{s_{\theta_{{\mathrm{x}}}} s_{\theta_{{\mathrm{y}}}}}{c_{\theta_{{\mathrm{y}}}}}  &\frac{c_{\theta_{{\mathrm{x}}}}s_{\theta_{{\mathrm{y}}}}}{c_{\theta_{{\mathrm{y}}}}}   \\
	 	0 & c_{\theta_{{\mathrm{x}}}} & -s_{\theta_{{\mathrm{x}}}}\\
	 	0 & \frac{s_{\theta_{{\mathrm{x}}}} }{c_{\theta_{{\mathrm{y}}}}} & \frac{c_{\theta_{{\mathrm{x}}}} }{c_{\theta_{{\mathrm{y}}}}}
	 \end{array}
	 \right]\left[\begin{array}{c}
		\omega_x \\
		\omega_y \\
     	\omega_z
	\end{array}\right]
\end{equation}
where $[\boldsymbol{\omega}_\mathrm{p}]_p=\left[\omega_x~\omega_y~\omega_z\right]^{\mathrm{T}}$ indicates the angular velocity of the platform in the frame $ \mathcal{F}_p$.
Further, the kinetic energy (KE)  and potential energy (PE) of the floating platform are
\begin{equation}\label{eq-4}
		\left\{
	\begin{aligned}
		&	\mathcal{T}_p=\frac{1}{2} m_p  [\mathbf{v}^{\mathrm{T}}_\mathrm{p}]_p [\mathbf{v}_\mathrm{p}]_p+\frac{1}{2}  [\boldsymbol{\omega}^{\mathrm{T}}_\mathrm{p}]_p\left(\mathbf{I}_p[\boldsymbol{\omega}_\mathrm{p}]_p\right),\\
		&	\mathcal{V}_{p}=m_p g r_z
	\end{aligned}
	\right.
\end{equation}
Furthermore, as depicted in Fig. \ref{fig0}, the position vectors in frame $\mathcal{F}_p$ for the centers of mass  of the tower ($G_t$), nacelle ($G_{nc}$), and rotor ($G_r$) are described by
\begin{equation*}
	\begin{aligned}
		 &
		 [\mathbf{r}_{G_t}]_p = \begin{bmatrix}
			0&0& H_{t}
		\end{bmatrix}^{\mathrm{T}},\quad
		[\mathbf{r}_{G_{nc}}]_p =\begin{bmatrix}
			h_{nc}&0& H_r
		\end{bmatrix}^{\mathrm{T}}\\
		 &[\mathbf{r}_{G_r}]_p =\begin{bmatrix}
			-h_r&0&H_r
		\end{bmatrix}^{\mathrm{T}}
	\end{aligned}
\end{equation*}
Meanwhile, the  velocities of these center of masses can be formulated by
\begin{equation*}
	\begin{aligned}
		&\mathbf{v}_t=\mathbf{v}_p+\boldsymbol{\omega}_p \times \mathbf{r}_{G_t},\quad \mathbf{v}_{n c}=\mathbf{v}_p+\boldsymbol{\omega}_p \times \mathbf{r}_{G_{n c}}  \\
		&\mathbf{v}_r=\mathbf{v}_p+\boldsymbol{\omega}_p \times \mathbf{r}_{G_r}
	\end{aligned}
\end{equation*}
Moreover, it is evident that the platform, tower, and nacelle share a common angular velocity denoted as $\boldsymbol{\omega}_p$, while the rotor's angular velocity is  $\boldsymbol{\omega}_r=\boldsymbol{\omega}_p+\boldsymbol{\Omega}_r$ and
the rotor spin rate is $[\boldsymbol{\Omega}_r]_p=\Omega_r\cdot[\mathbf{\hat{e}}_1]_p$.
Subsequently, the KE and PE of the tower, the nacelle, and the rotor are derived, respectively, denoted by:
\begin{equation}\label{eq-5}
	\begin{aligned}
		&\begin{cases}
			&\mathcal{T}_t = \frac{1}{2} m_t [\mathbf{v}^{\mathrm{T}}_\mathrm{t}]_p [\mathbf{v}_\mathrm{t}]_p + \frac{1}{2} [\boldsymbol{\omega}^{\mathrm{T}}_\mathrm{t}]_p \left(\mathbf{I}_t[\boldsymbol{\omega}_\mathrm{t}]_p\right),\\
			&\mathcal{V}_{t} = m_t g\left([\mathbf{r}_{to}]_I \cdot [\mathbf{\hat{e}}_3]_I\right)
		\end{cases}\\
	&	\begin{cases}
			&\mathcal{T}_\mathrm{nc} = \frac{1}{2} m_{nc} [\mathbf{v}^{\mathrm{T}}_\mathrm{nc}]_p [\mathbf{v}_\mathrm{nc}]_p + \frac{1}{2} [\boldsymbol{\omega}^{\mathrm{T}}_\mathrm{nc}]_p \left(\mathbf{I}_{nc}[\boldsymbol{\omega}_\mathrm{nc}]_p\right),\\
			&\mathcal{V}_\mathrm{nc} = m_t g\left([\mathbf{r}_{no}]_I \cdot [\mathbf{\hat{e}}_3]_I\right)
		\end{cases}\\
	&	\begin{cases}
			&\mathcal{T}_\mathrm{r} = \frac{1}{2} m_{r} [\mathbf{v}^{\mathrm{T}}_\mathrm{r}]_p [\mathbf{v}_\mathrm{r}]_p + \frac{1}{2} [\boldsymbol{\omega}^{\mathrm{T}}_\mathrm{r}]_p \left(\mathbf{I}_{r}[\boldsymbol{\omega}_\mathrm{r}]_p\right),\\
			&\mathcal{V}_\mathrm{r} = m_r g\left([\mathbf{r}_{r}]_I \cdot [\mathbf{\hat{e}}_3]_I\right)
		\end{cases}
	\end{aligned}
\end{equation}
with
\begin{equation*}
	\begin{aligned}
	&	[\mathbf{r}_{to}]_I=\mathbf{r}_p+\mathbf{R}[\mathbf{r}_{G_t}]_p,\quad [\mathbf{r}_{no}]_I=\mathbf{r}_p+\mathbf{R}[\mathbf{r}_{G_{nc}}]_p,\\
	&[\mathbf{r}_{ro}]_I=\mathbf{r}_p+\mathbf{R}[\mathbf{r}_{G_{r}}]_p
	\end{aligned}
\end{equation*}
Then the total KE and PE of the multibody system  can be obtained by
\begin{equation}\label{eq-6}
	\begin{aligned}
		\mathcal{T}=  &\mathcal{T}_p+\mathcal{T}_t+\mathcal{T}_{n c}+\mathcal{T}_r\\
		\mathcal{V}=&\mathcal{V}_{p}+\mathcal{V}_{t}+\mathcal{V}_{nc}+\mathcal{V}_{r}
	\end{aligned}
\end{equation}

\subsection{External forces}
The total external forces and accompanying moments exerted on the SSFWT primarily originate from hydrostatic, hydrodynamic, aerodynamic, and mooring loads, derived by
\begin{equation}\label{eq-7}
	\begin{aligned}
		\mathbf{F} & =\mathbf{F}_{B}+\mathbf{F}_{D}+\mathbf{F}_{A}+\mathbf{F}_{M} \\
		\mathbf{T} & =\mathbf{T}_{B}+\mathbf{T}_{D}+\mathbf{T}_{A}+\mathbf{T}_{M}
	\end{aligned}
\end{equation}
The subscripts $B,~D,~A~\text{and}~M$ correspond to the external forces and moments mentioned above in the description.

\begin{figure}[!htb]
	\begin{center}
		\centerline{\includegraphics[width=0.40\textwidth]{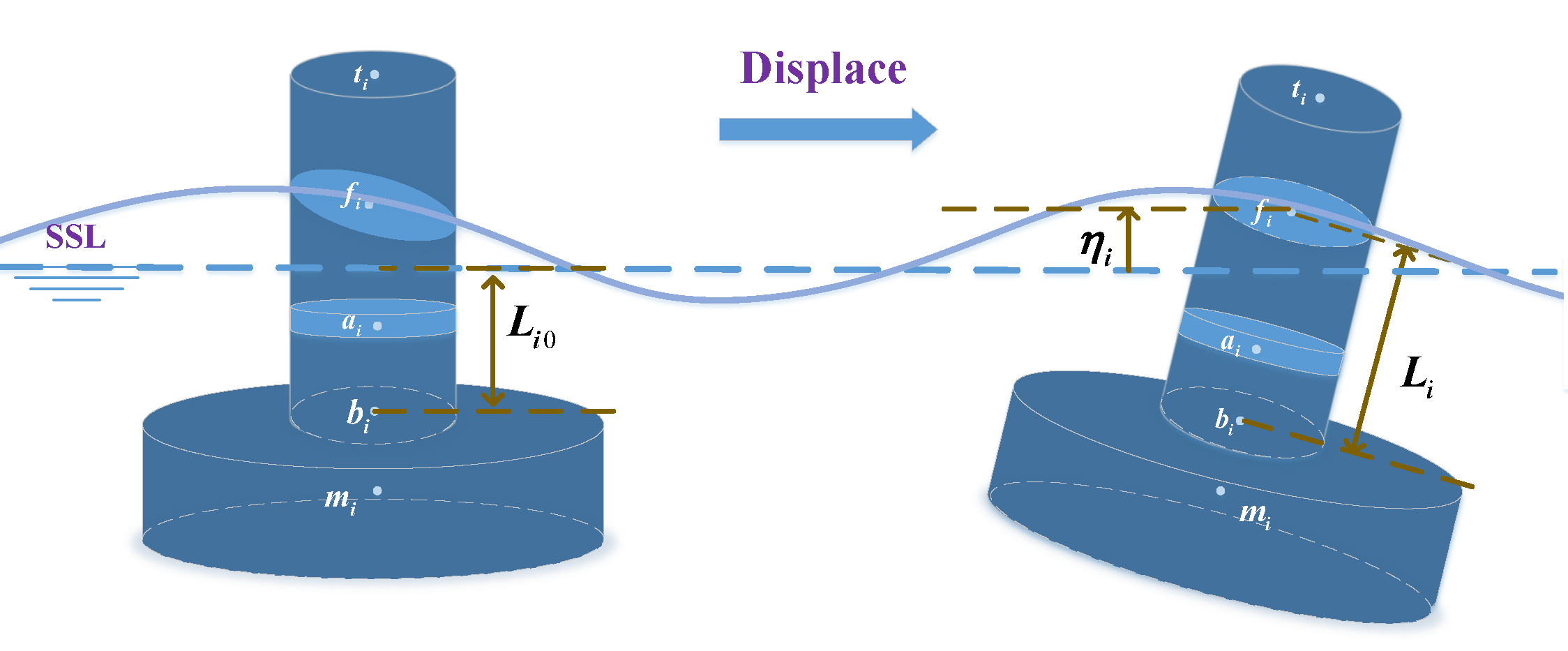}}
		\caption{Submerged length of platform floating member. }
		\label{fig1}
	\end{center}
\end{figure}

In this study, we assume that the semi-submersible platform comprises seven homogenous and uniform cylinders, denoted as a main column $(i = 1)$ and three upper side columns $(i = 2, 3, 4)$ floating in the water, along with three base columns $(i = 5, 6, 7)$ fully submerged.
As per Archimedes' principle, the buoyant force exerted on the platform equals the weight of the fluid displaced by the platform, as determined by
\begin{eqnarray}\label{eq-8}
		[\mathbf{F}_{\mathrm{B}}]_p=\mathbf{R}^{-1} \left((\rho_w g V_{d}) [\mathbf{\hat{e}}_3]_I \right)
\end{eqnarray}
where $ \rho_w$ and $g$ are the density of water and the gravitational constant, respectively; $V_{d}$ denotes the total instantaneous submerged volume, satisfying
\begin{equation}\label{eq-9}
	V_{d}=\sum_{i=1}^7 V_i, ~V_i=L_i A_i=\frac{\pi}{4} L_i d_{i}^2,~i=1,\cdots,7
\end{equation}
where $A_i$, $d_{i}$ and $L_i$ are the cross-sectional area,  the diameter  and  length of the $i$th cylinder, respectively.
In fact,  the length $L_i$ of the $i$th submerged members equals its actual length. For floating members, the magnitude of vector $[\mathbf{r}_{bf,i}]_I$, representing the vector from the base center  $b_i$  to the floating area center $f_i$ (as depicted in Fig. \ref{fig1}), characterizes its length.
Moreover,  the submerged length $L_i$  satisfies
\begin{equation}\label{eq-10}
\begin{aligned}
	&[\mathbf{r}_{bf,i}]_I=L_i\cdot \mathbf{R}[\mathbf{\hat{e}}_3]_p,\\
	&[\mathbf{r}_{bf,i}]_I\cdot[\mathbf{\hat{e}}_3]_I=\eta_i+L_{i0}+h_{bi}
\end{aligned}	
\end{equation}
which leads to
\begin{equation}\label{eq-11}
	\begin{aligned}
	     L_i=\frac{\eta_i+L_{i,0}+h_{b,i}}{\mathbf{R}[\mathbf{\hat{e}}_3]_p\cdot[\mathbf{\hat{e}}_3]_I}
	\end{aligned}
\end{equation}
where  the water free surface elevation, denoted as $\eta_i$, is the vertical displacement of the water surface as a wave traverses the $i$th floating member.
$L_{i 0}$ is the initial submerged length of the cylinder;
$h_{bi}$ denotes the vertical displacement  of each floating member's base center $b_i$,   described as
\begin{equation}\label{eq-12}
	\begin{aligned}
			h_{bi}&=\left(\mathbf{R}(0)[\mathbf{r}_{bi}]_p-([\mathbf{r}_p]_I+\mathbf{R}(\mathbf{\boldsymbol{\theta}_\mathrm{p}})[\mathbf{r}_{bi}]_p)\right)\cdot [\mathbf{\hat{e}}_3]_I\\
			&=(\mathbf{R}(0)-\mathbf{R}(\mathbf{\boldsymbol{\theta}_\mathrm{p}}))[\mathbf{r}_{bi}]_p\cdot[\mathbf{\hat{e}}_3]_I-r_z
	\end{aligned}
\end{equation}
where $[\mathbf{r}_{bi}]_p$ represents a constant vector in $\mathcal{F}_p$, originating from point $G_{p}$ and terminating at point $b_i$.
Furthermore,  the hydrostatic moment can be derived as
\begin{equation}\label{eq-13}
	\mathbf{T}_{\mathrm{B,i}}^p=[\mathbf{r}_{B}]_p \times [\mathbf{F}_{\mathrm{B}}]_p
\end{equation}
where $[\mathbf{r}_{B}]_p $ is the position vector of platform's buoyancy center in $\mathcal{F}_p$, which can be obtained by
 \begin{equation}\label{eq-14}
 	[\mathbf{r}_{B}]_p=\frac{\sum_{i=1}^7 V_i\cdot [\mathbf{r}_{mi}]_p}{V_d}
 \end{equation}
with $[\mathbf{r}_{mi}]_p=[\mathbf{r}_{bi}]_p+\frac{1}{2}L_i[\mathbf{\hat{e}}_3]_p,i=1,\ldots,7,$ being the position vector of
  centroid $m_i$ in  $\mathcal{F}_p$.

The hydrodynamic loads acting on the platform are determined utilizing the Morison equation and strip theory \cite{faltinsen1993sea}. Initially, the hydrodynamic force acting on a submerged cylinder in transverse water flow is calculated by
\begin{equation}\label{eq-15}
	\begin{aligned}
     \mathbf{F}_{\mathrm{dt,i}}=\mathbf{F}^t_{1i}+\mathbf{F}^t_{2i},~~i=1,\cdots, 7		
	\end{aligned}
\end{equation}
where $\mathbf{F}^t_{2i}$ denotes the inertial terms, $\mathbf{F}^t_{1i}$ represent the remaining terms of Morison's equation,  and they can be obtained by
\begin{equation*}
	\begin{aligned}
		\mathbf{F}^t_{1i}=&\int_{z_t}^{z_b} C_{Di} \|(\mathbf{v}_{ref,i})_{\perp}\| (\mathbf{v}_{ref,i})_{\perp}+(C_{Ai}+\rho_wA_i)\\
&\quad\cdot(\dot{\mathbf{v}}_{w,i}^p)_{\perp} dz,\\
		\mathbf{F}^t_{2i}=&-\int_{z_t}^{z_b} C_{Ai}(\dot{\mathbf{v}}_{a,i}^p)_{\perp}dz
	\end{aligned}
\end{equation*}
with $z_t=L_i,~z_b=0,~\mathbf{v}_{ref,i}=\mathbf{v}_{w,i}-\mathbf{v}_{a,i}$, and $C_{Di}=\frac{1}{2}C_{di}\rho_w d_i,~C_{Ai}=C_{ai}\rho_w A_i$.
Here,  $C_{d_{i}}$ and $C_{a_{i}}$ are drag and added mass coefficients, respectively;
$z_t$ and $z_b$ are the vertical coordinates of point $f_i$ and point  $b_i$ in $ \mathcal{F}_p$, respectively;
The amplitude of $(\cdot)$ is denoted by $\|(\cdot)\| $; The perpendicular component of $(\cdot)$ with respect to the centerline of the submerged member is denoted as $(\cdot)_{\perp}$.
Examining a length element $dz$ along each submerged platform member and a specific point $a_i$ situated on the centerline of the ith submerged segment, positioned at a distance of $z$ from point $b_i$, aligns with the centroid of the volume element, as shown in Fig. \ref{fig1}.
Then the position vector of $a_i$ in $\mathcal{F}_p$ is $[\mathbf{r}_{ai}]_p=[\mathbf{r}_{bi}]_p+z\cdot[\mathbf{\hat{e}}_3]_p $, and the velocity and acceleration  of $a_i$ are
\begin{equation*}
	\begin{aligned}
		 \mathbf{v}_{a,i}=\mathbf{v}_{p}+\mathbf{\omega}_{p}\times [\mathbf{r}_{ai}]_p,\quad \dot{\mathbf{v}}_{a,i}=\dot{\mathbf{v}}_{p}+\dot{\boldsymbol{\omega}}_{p}\times [\mathbf{r}_{ai}]_p
	\end{aligned}
\end{equation*}
Besides,   $\mathbf{v}_{w,i}$ and $\dot{\mathbf{v}}_{w,i}$ denote  the corresponding velocity and acceleration of water particle coincident with points $a_i$ in each moment.
In this study, the kinematics of the wave field is established through the application of the Airy wave theory. Comprehensive calculation details can be found in \cite{goda2010random}, and for brevity, they are not expounded upon in this context.

Then from \eqref{eq-15},  the moments caused by the hydrodynamic forces can be derived by
\begin{equation}\label{eq-16}
	\begin{aligned}
	\mathbf{T}_{\mathrm{dt,i}}= \mathbf{T}^t_{1i}+\mathbf{T}^t_{2i}  = \int_{z_t}^{z_b} \mathbf{r}_{a_i}\times \mathbf{F}_{\mathrm{dt,i}},~i=1,\cdots, 7		
	\end{aligned}		
\end{equation}
Moreover, it is essential to highlight that the three base columns of the platform ($i=5,6,7$) also experience drag and inertial forces in the heave direction. Similar to \eqref{eq-15}, these forces can be expressed as:
\begin{equation}\label{eq-17}
		\begin{aligned}
		&\mathbf{F}_{\mathrm{dh,i}}=\mathbf{F}^h_{1i}+\mathbf{F}^h_{2i}, ~i=5,6, 7		
	\end{aligned}
\end{equation}
where
\begin{equation*}
	\begin{aligned}
		\mathbf{F}^h_{1i}=& C_{D_zi} \|(\mathbf{v}_{ref,i})_{\|} \| (\mathbf{v}_{ref,i})_{\|}+C_{A_zi}(\dot{\mathbf{v}}_{w,i}^p)_{\|} dz,\\
		&+(A_ip_{bi}-(A_i-A_{i-3})p_{ti})\cdot[\mathbf{\hat{e}}_3]_p,\\
		\mathbf{F}^h_{2i}=&-C_{A_zi}(\dot{\mathbf{v}}_{c_i})_{\|}
	\end{aligned}
\end{equation*}
with $C_{D_zi}=\frac{1}{2}C_{d_zi}\rho_w A_i$, $C_{A_zi}=C_{a_zi}\rho_w V_{Ri}$ and $V_{\mathrm{Ri}}=\frac{2}{3} \pi ( d_i/2  )^3 $,
where  $C_{\mathrm{dz,i}}$ and $C_{\mathrm{az,i}}$ represent the drag and added mass coefficients along the $Z$-axis, respectively;  $p_{bi}$ and $p_{ti}$ denote the dynamic pressure at the bottom and top faces of the heave-plate. The  symbol  $(\cdot)_{\|}$  indicates a component parallel to the centerline of the submerged member.
Point $c_i$ corresponds to the center of gravity for the $i$th base column, with its velocity and acceleration given by:
\[
\mathbf{v}_{c_i}=\left.\mathbf{v}_{\mathrm{a}_i}\right|_{z=\frac{L_i}{2}},~~\dot{\mathbf{v}}_{c_i}=\left.\dot{\mathbf{v}}_{\mathrm{a}_i}\right|_{z=\frac{L_i}{2}}
\]
Correspondingly,  the moments caused by the hydrodynamic forces can be derived by
\begin{equation}\label{eq-18}
	\begin{aligned}
		\mathbf{T}_{\mathrm{dh,i}}=&\mathbf{T}^h_{1i}+\mathbf{T}^h_{2i}
		= \mathbf{r}_{c_i}\times(\mathbf{F}^h_{1i}+\mathbf{F}^h_{2i}),~i=5,6, 7		
	\end{aligned}		
\end{equation}
with $ \mathbf{r}_{c_i}=\left.\mathbf{r}_{\mathrm{a}_i}\right|_{z=\frac{L_i}{2}} $.

To sum up, the total hydrodynamic force and corresponding moments on the platform are
\begin{equation}\label{eq-19}
	\begin{aligned}
	&\mathbf{F}_{\mathrm{D}}=\sum_{i=1}^{7} \mathbf{F}_{\mathrm{dt,i}}+\sum_{j=5}^{7} \mathbf{F}_{\mathrm{dh,j}},\\
		&\mathbf{T}_{\mathrm{D}}=\sum_{i=1}^{7} \mathbf{T}_{\mathrm{dt,i}}+\sum_{j=5}^{7} \mathbf{T}_{\mathrm{dh,j}}
	\end{aligned}
\end{equation}
\begin{remark}
	It is noteworthy that the inertial terms in \eqref{eq-15}--\eqref{eq-18}, namely, $\mathbf{F}^t_{2i}$, $\mathbf{F}^h_{2i}$, $\mathbf{T}^t_{2i}$, $\mathbf{T}^h_{2i}$, are typically relocated to the left-hand side of the equation of motion.
	 This results in the formation of an added mass matrix $M_{a}$, and its formulation  is provided in Appendix \ref{app1}.
\end{remark}

Next, the primary wind-induced loads on the SSFWT involve the aerodynamic torque $\tau_a$ around the turbine shaft and the axial thrust force $F_a$ along the shaft axis, expressed as
\begin{equation}
	\begin{aligned}
		F_{a} & =\frac{1}{2} \rho_{a} \pi R_{r}^2 C_t(\lambda, \beta)\|(\mathbf{u}_{w} )_{\|}\|^2 \\
		\tau_{a} & =\frac{1}{2} \rho_{a} \pi R_{r}^3 \frac{1}{\lambda} C_p(\lambda, \beta)\|(\mathbf{u}_{w})_{\|}\|^2
	\end{aligned}
\end{equation}
where $R_{r}$ signifies the rotor radius, while $\rho_{a}$ is air density. The relative velocity between the wind and rotor is denoted by $\mathbf{u}_{w}$, which can be calculated as $\mathbf{u}_{w}=\mathbf{v}_w-\mathbf{v}_r$,
where $\mathbf{v}_w$ is the wind speed vector. Further, we can obtain that the  component of $\mathbf{u}_{w}$ and parallel to the rotor shaft axis $([\mathbf{\hat{e}}_1]_p)$ can be derived by
$
(\mathbf{u}_{w})_{\|}=(\mathbf{u}_{w}\cdot[\mathbf{\hat{e}}_1]_p )[\mathbf{\hat{e}}_1]_p.
$
The thrust and power coefficients of  WTs are denoted as $C_t$ and $C_p$, respectively, which are dependent on the blade pitch angle $\beta$ and the tip speed ratio $\lambda$, with $\lambda= \Omega_r R_{r}/ \|(\mathbf{u}_{w})_{\|}  \|  $.
Then the formulations of the aerodynamic force and associated moment around the origin of frame $\mathcal{F}_p$ are  as follows
\begin{equation}
	[\mathbf{F}_{A}]_p  = F_a [\mathbf{\hat{e}}_1]_p,~ [\mathbf{T}_{A}]_p  =\tau_a [\mathbf{\hat{e}}_1]_p+   [\mathbf{r}_{r}]_p  \times [\mathbf{F}_{A}]_p
\end{equation}

In addition, mooring loads are evaluated utilizing the quasi-static method \cite{irvine1981cable,isaacson1996moored}. For the detailed calculation process,  readers can refer to \cite{al2019dynamic}, while a thorough discussion is omitted here to maintain brevity.

\subsection{Nonlinear model of SSFWT}
In order to obtain the dynamic model of SSFWT,
the Lagrange's equation based on the platform's quasi-coordinates ($\mathbf{v}_\mathrm{p}, \boldsymbol{\omega}_\mathrm{p}$)  is  utilized to derives the equations of motion (EOM) of the platform, while the EOM  of the WT drivetrain is obtained  from the conventional form of Language' equation \cite{meirovitch1995hybrid,meirovitch2003integrated}.
By recalling equations \eqref{eq-2}, \eqref{eq-3}, \eqref{eq-6}, and \eqref{eq-7}, the nonlinear model of SSFWT can be derived as:
	\begin{equation}\label{eq-23}
		\mathbf{\bar{M}} \begin{bmatrix}
			[\dot{\mathbf{r}}_\mathrm{p}]_I\\
			[\dot{\boldsymbol{\theta}}_\mathrm{p}]_I\\
			[\dot{\mathbf{v}}_\mathrm{p}]_p\\
			[\dot{\boldsymbol{\omega}}_\mathrm{p}]_p\\
			\dot{\Omega_r}
		\end{bmatrix}=\begin{bmatrix}
			\mathbf{R}[\mathbf{v}_\mathrm{p}]_p\\
			\mathbf{J}[\boldsymbol{\omega}_\mathrm{p}]_p\\
			 [\mathbf{F}]_p  \\
			 [\mathbf{T}]_p \\
			\tau_a-\tau_g
		\end{bmatrix}+ \mathbf{s}(\mathbf{r}_\mathrm{p}, \boldsymbol{\theta}_\mathrm{p}, \mathbf{v}_\mathrm{p}, \boldsymbol{\omega}_\mathrm{p}, \Omega_r)
	\end{equation}
where $\mathbf{s}$ is a vector function about $\mathbf{r}_\mathrm{p}, \boldsymbol{\theta}_\mathrm{p}, \mathbf{v}_\mathrm{p}, \boldsymbol{\omega}_\mathrm{p}, \Omega_r$.
The expressions of $\mathbf{\bar{M}} $ and $\mathbf{s} $ are given in Appendix \ref{app2}.
\begin{remark}
 Note that the simplified SSFWT model developed in \eqref{eq-23} is primarily utilized for controller design, with subsequent  validation typically conducted on the complete state model.
\end{remark}

\section{Controller design}
In this section,  a novel  pitch controller for SSFWT is designed based on the robust integral of the sign of the error (RISE) control approach, developed in \cite{xian2004continuous}, and then
 the  regulation of generator power while reducing the platform motion can be guaranteed in  region III.

\subsection{Model reformulation}
Based on \eqref{eq-23}, the SSFWT model can be rewritten as
\begin{equation}\label{eq-24}
	\dot{\mathbf{X}}= \mathbf{\bar{M}}^{-1}(\boldsymbol{\Lambda}(\mathbf{X},\beta,\mathbf{v}_w)+\mathbf{s}(\mathbf{X}))
\end{equation}
with $\mathbf{X}=\left[[\mathbf{r}^T_\mathrm{p}]_I~[\boldsymbol{\theta}^T_\mathrm{p}]_I~[\mathbf{v}^T_\mathrm{p}]_p~[\boldsymbol{\omega}^T_\mathrm{p}]_p~\Omega_r\right]^T$ and
\begin{equation*}
	\boldsymbol{\Lambda}^T=\left[(\mathbf{R}[\mathbf{v}_\mathrm{p}]_p)^T ~~(\mathbf{J} [\boldsymbol{\omega}_\mathrm{p}]_p)^T~~ [\mathbf{F}]_p^T~ ~[\mathbf{T}]_p^T~~\tau_a-\tau_g\right]
\end{equation*}
For the purpose of controller design, the external load vector $\boldsymbol{\Lambda}$ is divided as $\boldsymbol{\Lambda}(\mathbf{X},\beta,\mathbf{v}_w) =\boldsymbol{\Lambda}_a(\mathbf{X},\beta,\mathbf{v}_w) +\boldsymbol{\Lambda}_e(\mathbf{X}) $, with
\begin{equation}\label{eq-25}
	\begin{aligned}
		\boldsymbol{\Lambda}_a(\mathbf{X},\beta,\mathbf{v}_w)=&[[\mathbf{F}^{\mathrm{T}}_{A}]_p~~  [\mathbf{T}^{\mathrm{T}}_{A}]_p~~ \tau_a]^{\mathrm{T}}
	\end{aligned}
\end{equation}
 where $\boldsymbol{\Lambda}_a$ is a vector related to pitch angle $\beta$.
Then we have
\begin{equation}\label{eq-26}
	\begin{aligned}
		 	\dot{\mathbf{X}}=f_0(\mathbf{X})+g_0(\mathbf{X},\beta,\mathbf{v}_w)+\mathbf{d}(t)
	\end{aligned}
\end{equation}
with
\begin{equation}\label{eq-27}
	\begin{aligned}
		&f_0(\mathbf{X})=\mathbf{\bar{M}}^{-1}(\mathbf{s}(\mathbf{X})+\Lambda_e(\mathbf{X}))\\
		&g_0(\mathbf{X},\beta,\mathbf{v}_w)=\mathbf{\bar{M}}^{-1} \Lambda_a(\mathbf{X},\beta,\mathbf{v}_w)
	\end{aligned}
\end{equation}
and $\mathbf{d}(t)$ includes the uncertain parameters of model and external disturbances, which is assumed to satisfy: $\|d(t)\| \leq d_{m 0}$,
$\|\dot{d}(t)\| \leq d_{m 1}$, and $\|\ddot{d}(t)\| \leq d_{m 2}$, where $d_{m 0}, d_{m 1}, d_{m 2}$ are unknown positive constants.

\subsection{Problem formulation }
In SSFWTs, maintaining generator power at its rated value corresponds to regulating the rotor speed  $\Omega_r$  to its rated value $\Omega_0$, denoted as $\Omega_r \rightarrow \Omega_0$. Similarly, minimizing platform pitch motion involves reducing the platform pitch rate, represented as $\dot{\theta}_y \rightarrow 0$. Unlike onshore wind turbines, SSFWTs may experience negative damping phenomena \cite{tran2015platform} when the relative wind speed permits the rotor to rotate at its rated value $\Omega_{r0}$.
To address this concern, an efficient resolution, inspired by the findings in \cite{lackner2009controlling}, involves redefining the target rotor speed as a function of the platform pitch velocity $\dot{\theta}_y$, i.e.,
\begin{equation}\label{eq-28}
	\Omega^{*}_{r}=\Omega_{r0}-k \dot{\theta}_y,~k>0
\end{equation}
Then a trade-off between the aforementioned control objectives can be attained.
Besides, it should be noted that while the control objective in our study only focuses on two DOFs of FWT, i.e., the rotor speed and platform pitch motion, it is essential to acknowledge the FWT is a multi-body coupled system and there exists inherent interplay among various DOFs, as presented in \eqref{eq-23}. Consequently, in order to accurately capture the dynamic response of the system, we have opted to establish a comprehensive 7-DOFs mathematical model of the FWT.

\subsection{Pitch controller design }
Define $\xi(t)=\Delta\Omega_r+k\Delta \dot{\theta}_y$,
with
$
        \Delta\Omega_r= \Omega_r-\Omega_{r0},~\Delta \dot{\theta}_y= \dot{\theta}_y-0. $ Then from  \eqref{eq-3}, one has
\begin{equation}\label{eq-29}
	\begin{aligned}
		\xi(t)=\boldsymbol{m}^T\mathbf{X}-\Omega_{r0}
	\end{aligned}	
\end{equation}
with $\boldsymbol{m}=[\mathbf{0}_{1\times9}~0~k c_{\theta_{{\mathrm{x}}}} ~-ks_{\theta_{{\mathrm{x}}}}~1]^T$. Further, taking the time derivative of $\xi(t)$ gives
 \begin{equation}\label{eq-30}
 	\begin{aligned}
 		\dot{\xi}(t)&=\dot{\mathbf{m}}^T\mathbf{X}+\mathbf{m}^T\dot{\boldsymbol{X}}\\
 		&=\dot{\mathbf{m}}^T\mathbf{X}+\mathbf{m}^T(f_0(\mathbf{X})+g_0(\mathbf{X},\beta,\mathbf{v}_w)+\mathbf{d}(t) )
 	\end{aligned}	
 \end{equation}
Let
 	\begin{equation}\label{eq-31}
 	\bar{\xi}=\dot{\xi}+c\xi, ~c>0
 \end{equation}
Taking the second derivative of $\xi(t)$ obtains
\begin{equation}\label{eq-32}
	\begin{aligned}
		\ddot{\xi}=&\ddot{\mathbf{m}}^T\mathbf{X}+\dot{\mathbf{m}}^T\dot{\mathbf{X}}+\dot{\mathbf{m}}^T (f_0(\mathbf{X})+g_0(\mathbf{X},\beta,\mathbf{v}_w)\\
		&+\mathbf{d}(t) )+\mathbf{m}^T(\frac{\partial {f}_{0}}{\partial {\mathbf{X} }}+\frac{\partial {g}_{0}}{\partial {\mathbf{X} }})\dot{\mathbf{X}}+\mathbf{m}^T\frac{\partial {g}_{0}}{\partial {\beta}}\dot{\beta}\\
		&+\mathbf{m}^T\frac{\partial {g}_{0}}{\partial {v_w}}\dot{v}_w+\mathbf{m}^T\dot{\mathbf{d}}(t)
	\end{aligned}
\end{equation}
Moreover,  differentiating \eqref{eq-31} obtains
 \begin{equation}\label{eq-33}
 	\begin{aligned}
 		\dot{\bar{\xi}}= &\ddot{\xi}+c\dot{\xi}\\
 		=&f_1(\mathbf{X},\beta,\mathbf{v}_w)+g_1(\mathbf{X},\beta,\mathbf{v}_w)\dot{\beta}+d_1(t)
  	\end{aligned}	
 \end{equation}
with $
g_1(\mathbf{X},\beta,\mathbf{v}_w)=\mathbf{m}^T\frac{\partial {g}_{0}}{\partial {\beta}}$, and
\begin{equation*}
	\begin{aligned}
		f_1(\mathbf{X},\beta,\mathbf{v}_w)=&\ddot{\mathbf{m}}^T\mathbf{X}+2\dot{\mathbf{m}}^T (f_0(\mathbf{X})+g_0(\mathbf{X},\beta,\mathbf{v}_w))\\
		&+\mathbf{m}^T(\frac{\partial {f}_{0}}{\partial {\mathbf{X} }}+\frac{\partial {g}_{0}}{\partial {\mathbf{X} }})(f_0(\mathbf{X})+g_0(\mathbf{X},\beta,\mathbf{v}_w) )\\
	    d_1(t)=&2\dot{\mathbf{m}}^T\mathbf{d}(t) +\mathbf{m}^T(\frac{\partial {f}_{0}}{\partial {\mathbf{X} }}+\frac{\partial {g}_{0}}{\partial {\mathbf{X} }})\cdot\mathbf{d}(t)\\
	    &+\mathbf{m}^T\frac{\partial {g}_{0}}{\partial {v_w}}\dot{v}_w+\mathbf{m}^T\dot{\mathbf{d}}(t)
	\end{aligned}
\end{equation*}
Combining \eqref{eq-25} and \eqref{eq-27},  it can be derived that
\begin{equation}\label{eq-34}
	\begin{aligned}
	g_1(\mathbf{X},\beta,\mathbf{v}_w)&=\mathbf{m}^T \mathbf{\bar{M}}^{-1}\frac{\partial}{\partial {\beta}}  \Lambda_a(\mathbf{X},\beta,\mathbf{v}_w)\\
	&=b_1\frac{\partial \tau_a}{\partial {\beta}} +b_2kc_{\theta_{x}}\frac{\partial F_a}{\partial {\beta}}
	\end{aligned}
\end{equation}
where $b_1$ and $b_2$ are both positive constant, whose specific mathematical expressions are given in the Appendix \ref{app2}.
  Since $\frac{\partial {\tau}_{a}}{\partial {\beta}}<0,~ \frac{\partial {F}_{a}}{\partial {\beta}}<0$. Thus, one has
  \[
   g(\mathbf{X},\beta,\mathbf{v}_w)<0
  \]
Define $T(\mathbf{X},\beta,\mathbf{v}_w)=- 1/g(\mathbf{X},\beta,\mathbf{v}_w)  $. In fact,  $T(\mathbf{X},\beta,\mathbf{v}_w)$ is a positive scalar with $ 1/g_{\min}$ being its largest singular value, where $g_{\min}$ is the minimum value of function $-g(\mathbf{X},\beta,\mathbf{v}_w)$.
By utilizing \eqref{eq-33}, one has
\begin{equation}\label{eq-36}
	\begin{aligned}
		T(\mathbf{X},\beta,\mathbf{v}_w)\dot{\bar{\xi}}=&-\frac{1}{2} \dot{T}(\mathbf{X},\beta,\mathbf{v}_w) \bar{\xi}-\xi-\dot{\beta}\\
		&+F(\mathbf{X},\beta,\mathbf{v}_w)+D(t)
	\end{aligned}
\end{equation}
with
\begin{equation}\label{eq-37}
	\begin{aligned}
		F(\mathbf{X},\beta,\mathbf{v}_w)=&	-\frac{f(\mathbf{X},\beta,\mathbf{v}_w)}{g(\mathbf{X},\beta,\mathbf{v}_w) }+\frac{1}{2} \dot{T}(\mathbf{X},\beta,\mathbf{v}_w) \bar{\xi}+\xi\\
		D(t)=&T(\mathbf{X},\beta,\mathbf{v}_w) d_1(t)
	\end{aligned}
\end{equation}
\begin{remark}
Utilizing the mean value theorem \cite{malik1992mathematical}, we deduce the boundedness of the functions $D(t)$ and $\dot{D}(t)$, expressed as $|D(t)| \leq D_{m 0}$ and $|\dot{D}(t)| \leq D_{m 1}$, respectively. Here, $D_{m 0}$ and $D_{m 1}$ denote unknown positive constants.
\end{remark}
Notably, the uncertain dynamics  is divided into two parts in \eqref{eq-36}, i.e.,  $F(\mathbf{X},\beta,\mathbf{v}_w)$  and $D(t)$, which means that  the disturbance signal $\mathbf{d}(t)$ is absent from the function $F(\mathbf{X}, \beta, \mathbf{v}_w)$. Thus, an  online approximator (OLA) can be utilized to approximate and compensate $F(\mathbf{X},\beta,\mathbf{v}_w)$. Specifically, there exists an ideal neural network (NN) approximation such that
 \begin{equation}\label{eq-38}
 		F(\mathbf{X},\beta,\mathbf{v}_w)=W^{* T} \phi( Z)+\epsilon( Z)
 \end{equation}
 where
 $Z=[\mathbf{X}^T,\beta,\mathbf{v}_w]^T$. $W^{*} \in \mathbb{R}^{N \times 1} $ is the  ideal weight matrix satisfying $ \|W^{*}\| \leq w_m$, and  $N$ is the number of hidden layer nodes.   $\phi(\cdot): \mathbb{R}^{N} \rightarrow \mathbb{R}^{N}$ is the activation function  satisfying $\|\phi\| \leq \phi_{m 0},~\|\dot{\phi}\| \leq \phi_{m 1} $; $\epsilon( Z)$
 indicate the bounded NN reconstruction error  satisfying $ \|\epsilon\| \leq \epsilon_{m0}$, $ \|\dot{\epsilon}\| \leq \epsilon_{m1}$.
 Note that the ideal weight matrix $W^{*}$ is usually not known and must be estimated online for controller design purposes.
 Let $\hat{W}$ represent the estimate of $W^{*}$.
Then, a robust adaptive  pitch controller is designed as
 \begin{equation}\label{eq-39}
 	\begin{aligned}
 		\dot{\beta}=(k_c+1) \bar{\xi}+\hat{W}^{\mathrm{T}} \phi(Z)+\gamma(t) \operatorname{sgn}(\xi)
 	\end{aligned}
 \end{equation}
which leads to
\begin{equation}\label{eq-40}
\begin{aligned}
	\beta= & \left(k_c+1\right) \xi(t)-\left(k_c+1\right) \xi(0)   +\int_0^t [\hat{W}^T \phi(Z(\varsigma )) \\
	&+\left(k_c+1\right) c \xi (\varsigma )+\gamma(\varsigma ) \operatorname{sgn}(\xi(\varsigma )) ] d \varsigma
\end{aligned}
\end{equation}
where $k_c>0$ represents a constant gain, $\gamma(t)$ (with its expression to be specified later) serves as an adaptive term, and $\operatorname{sgn}(\cdot)$ denotes the signum function.
 Moreover, the learning algorithm of $\hat{W}$ is given as
 \begin{equation}\label{eq-41}
 	\dot{\hat{W}}=l_w(c\phi(Z) \xi-k_w|\xi| \hat{W})
 \end{equation}
Based on \cite{he2015adaptive}, it can be deduced that $\|\hat{W}\| \leq c \phi_{m 0}/k_w$.

\begin{figure}[!htb]
    \centering
        \includegraphics[width=0.45\textwidth]{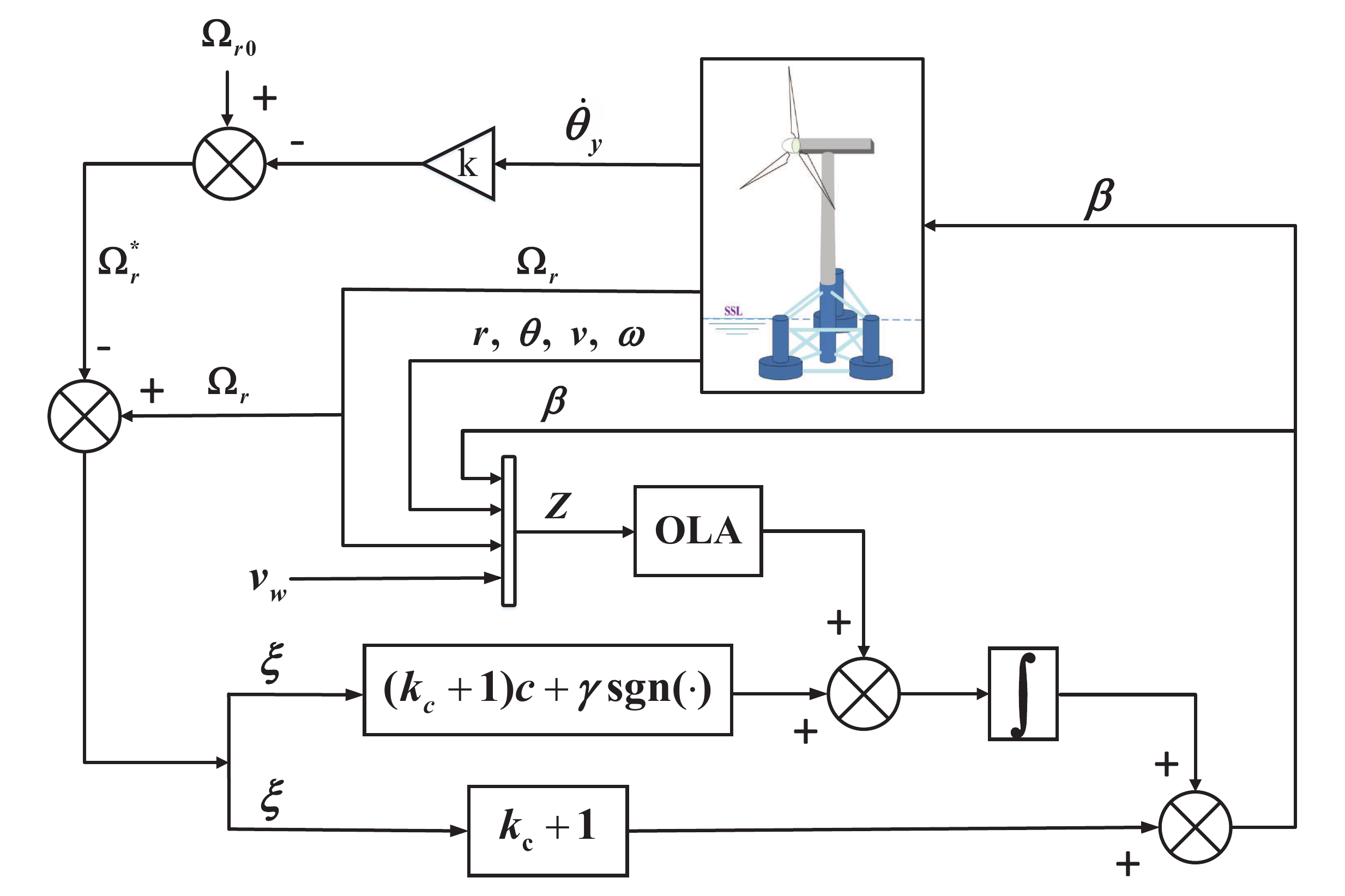}
    \caption{Overall structure of the proposed controller.}
    \label{fig-4}
\end{figure}

Next,   the adaptive term $ \gamma$ in \eqref{eq-39} can be designed as
\begin{equation}\label{eq-42}
	\begin{aligned}
		\gamma & =  \int_0^t \bar{\xi}^T(\varsigma ) \operatorname{sgn}(\xi(\varsigma )) d \varsigma  \\
		& =   \int_{\xi(0)}^{\xi(t)} \operatorname{sgn}\left(\xi\right) d \xi+  c \int_0^t|\xi(\varsigma )| d \xi
	\end{aligned}
\end{equation}
Without loss of generality, we assume
that $\xi$ changes its sign at a finite number of moments $t_i$, $i = 0, \ldots, m-1$, where $m > 1$. That is, $\xi(t_i) = 0$, and for all $p_1, p_2 \in [t_i, t_{i+1}]$, we have $\operatorname{sgn}(\xi(p_1)) = \operatorname{sgn}(\xi(p_2))$.
Let $t_0 = 0$ and $t_m = t$. Then $ \gamma$ can be reconstructed as
\begin{equation}
	\label{eq-43}
	\begin{aligned}
		\gamma & =  \int_{\xi(0)}^{\xi(t)} \operatorname{sgn}\left(\xi\right) d \xi+ c \int_0^t|\xi(\varsigma )| d \varsigma  \\
		& = \sum_{i=0}^{m-1} \int_{\xi\left(t_{i}\right)}^{\xi\left(t_{i+1}\right)} \operatorname{sgn}\left(\xi\right) d \xi+ c \int_0^t|\xi(\varsigma )| d \varsigma  \\
		& = \sum_{i=0}^{m-1}\left(\left|\xi \left(t_{i}\right)\right|-\left|\xi\left(t_{i}\right)\right|\right)+ c \int_0^t|\xi(\varsigma )| d \varsigma  \\
		& =|\xi(t)|-|\xi(0)|+ c \int_0^t|\xi(\varsigma )| d \varsigma
	\end{aligned}
\end{equation}
Further, taking the time derivative of $ \gamma$ gives
\begin{equation}
	\label{eq-44}
	\dot{\gamma}=\bar{\xi}^T\operatorname{sgn}(\xi)
\end{equation}
Finally, the overall structure  of the presented adaptive pitch controller   is shown Fig. \ref{fig-4}.

\subsection{Stability analysis}
Stability and performance analyses based on the standard Lyapunov method are conducted in the subsequent description.

First, substituting \eqref{eq-38} and \eqref{eq-39} into \eqref{eq-37} leads to
\begin{equation}
	\begin{aligned}
		T(\mathbf{X},\beta,\mathbf{v}_w)\dot{\bar{\xi}}=&-\frac{1}{2} \dot{T}\bar{\xi}-\xi-(k_c+1) \bar{\xi}\\
		&-\gamma(t) \operatorname{sgn}(\xi)+\tilde{W}^{\mathrm{T}} \phi(Z)+\bar{D}(t)
	\end{aligned}
\end{equation}
 where $\tilde{W}=W-\hat{W}, \bar{D}(t)=D(t)+\epsilon(Z)$.
 It is clear that $\|\bar{D}(t)\| \leq D_{m 0}+\varepsilon_{m 0} \equiv \bar{D}_{m 0}$ and $\|\dot{\bar{D}}(t)\| \leq D_{m 1}+\varepsilon_{m 1} \equiv \bar{D}_{m 1}$, and recalling \eqref{eq-24} and \eqref{eq-39}, we have
 \begin{equation}
 	 \begin{aligned}
 		& \left\|\tilde{W}^T \phi(Z) \right\| \leq W_{m 0}, \\
 		& \left\|\dot{\tilde{W}}^T \phi(Z)+\tilde{W}^T \dot{\phi}(Z) \dot{Z}\right\| \leq W_{m 1}
 	\end{aligned}
 \end{equation}
 where $W_{m 0}$ and $W_{m 1}$ are unknown positive  constants.
 Also, to support the stability analysis, the following lemma is given.
\begin{lemma}\cite{yang2015robust}
	Define
	\begin{equation*}
		\begin{aligned}
			&\gamma_d \equiv\bar{D}_{m 0}+\max \{ W_{m 0},\frac{(\bar{D}_{m 1}+W_{m 1}+k_w\|w_m\|^2/4 )}{c}\},\\
			&H\equiv\bar{\xi}^T\left(\bar{D}-\gamma_d \operatorname{sgn}(\xi)\right)+k_w|\xi| \operatorname{tr}(\tilde{W}^T \hat{W})-\dot{\xi}^T \bar{N}
		\end{aligned}	
	\end{equation*}
	with $\bar{N}=\tilde{W}^T \phi(Z)$. Then it can be concluded
	\begin{equation}
		\int_0^t H(\varsigma ) d \varsigma  \leq H_0
	\end{equation}
	with  $H_0\equiv \gamma_d|\xi(0)|-\xi^T(0)(\bar{D}(0)+\bar{N}(0))$.
\end{lemma}


In fact, the adaptive tuning law in \eqref{eq-44} is designed to facilitate the gradual convergence of the parameter $\gamma$ towards an unknown desired value $\gamma_d$. The approximation error is defined as $\tilde{\gamma}=\gamma_d-\gamma$. Subsequently, the stability  of blade pitch controller is presented in the following theorem.
\begin{theorem}
	Given the  nonlinear model of SSFWT in \eqref{eq-26}. The desired rotor speed is described in \eqref{eq-28}.
Then the pitch angle controller  in  \eqref{eq-40} with the adaptive laws \eqref{eq-41} and  \eqref{eq-42} can achieve the convergence of the regulation error.
\end{theorem}
\begin{IEEEproof}
	See the Appendix \ref{app3}.
\end{IEEEproof}

\section{Model validation and control results }


This section aims to assess the efficacy of the presented nonlinear modeling approach and the blade pitch controller for SSFWTs.
The simulation utilizes parameters outlined in \cite{jonkman2009definition, robertson2014definition} for the proposed model.

\subsection{Model validation}

\begin{figure}[!htb]
	\centering
	\includegraphics[width=0.45\textwidth]{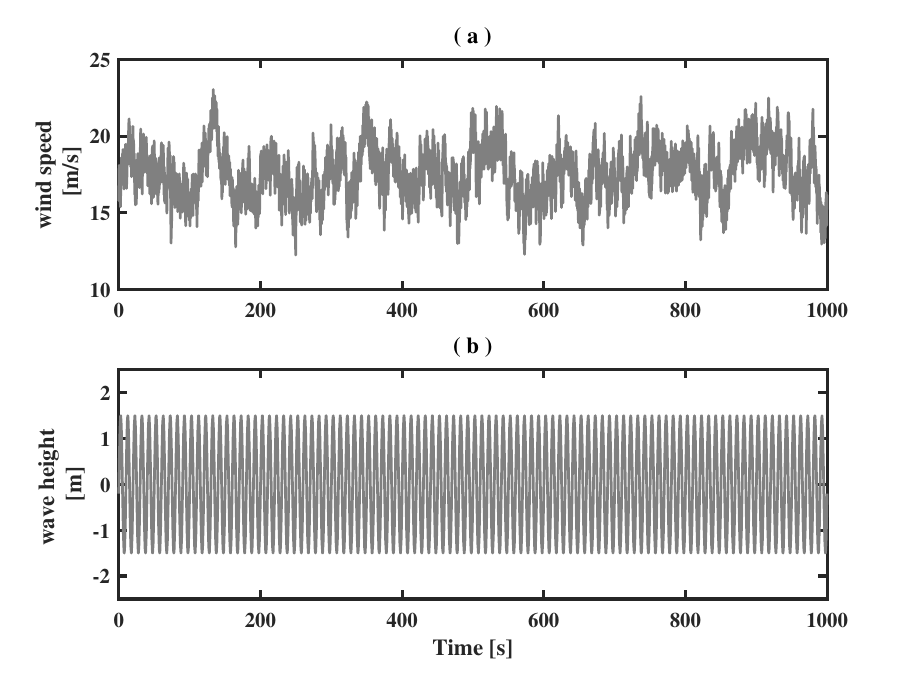}
	\caption{Profiles of wind and wave height.}
	\label{fig3}
\end{figure}

First, an open-loop response test is implemented for the validation of  the proposed SSFWT model under wind and wave joint loads.
In particular, the turbulent wind data features a mean hub-height wind speed of 18 m/s and a turbulence intensity of 10\%, with corresponding results depicted in the top of Fig. \ref{fig3}.
For regular harmonic wave conditions, characterized by a peak period, $T_p=10~s$, and a significant wave height of $H_s=3~m$,  the wave height time history is presented in the bottom of Fig. \ref{fig3}.
In addition, the simulation commences with initial displacement and angular conditions set at $5~m$ and $9^{\circ}$, respectively. The total simulation duration spans $1000s$, during which the blade pitch angle $\beta$ remains fixed at zero.
Moreover, the system responses generated by the presented nonlinear model are systematically compared with those derived from  the widely used software FAST  with all DOFs activated in the time domain.

\begin{figure}[!htb]
    \centering
        \includegraphics[width=0.45\textwidth]{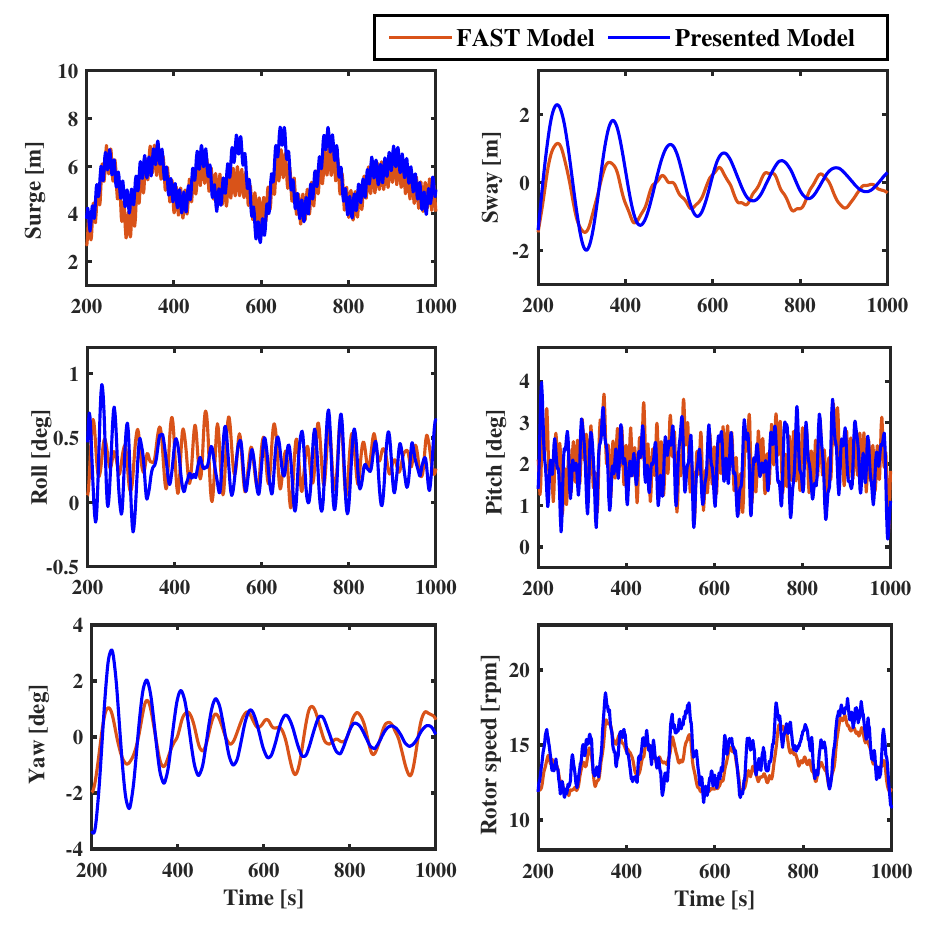}
    \caption{The system responses under wind-wave joint loads.}
    \label{fig4}
\end{figure}

Fig. \ref{fig4} presents a comparative analysis of the platform motion responses and rotor spin speed response for  SSFWT.
 To further substantiate the efficacy of the presented model, average (AV) values and root mean square (RMS) values of the responses are quantitatively assessed and presented in Table \ref{tab1}.
The findings conclusively demonstrate the proficiency of the proposed SSFWT model in accurately predicting the primary motions of the complete 44-state FAST model, while considering the interplay of coupled 3-D motion dynamics.
Notably, the responses in surge, sway, roll, and pitch  DOFs exhibit a qualitative resemblance between the FAST and the presented model. The identified disparities in these DOFs amount to 4.6\%, 2,2\%, 2.9\%, and 1.5\%, respectively, substantiating the precision of the proposed model in capturing the fundamental dynamic behaviors of the SSFWT system.

\begin{table}
	\small
	\centering
	\caption{Statistical characteristics of the system responses.}
	\renewcommand{\arraystretch}{1.2} %
	\begin{tabular}{
			l
			S[table-format=2.2]
			S[table-format=2.2]
		  c
		}
		\toprule\toprule
	\multirow{2}{*}{~}  & \multicolumn{1}{c}{\textbf{FAST}} & \multicolumn{1}{c}{\textbf{Presented}}& \multirow{2}{*}{\textbf{Discrepancy}}\\
    & \multicolumn{1}{c}{\textbf{Model}} & \multicolumn{1}{c}{\textbf{Model}} & \\
		\midrule
		$\operatorname{AV}(r_x)~[\mathrm{m}]$ & 5.11 & 5.42 & 5.63\% \\
		$\operatorname{RMS}(r_x)~[\mathrm{m}]$ & 5.18 & 5.50 & 5.94\% \\
		$\operatorname{AV}(\theta_y)~[\mathrm{deg}]$ & 2.10 & 1.99 & 5.31\% \\
		$\operatorname{RMS}(\theta_y)~[\mathrm{deg}]$ & 2.18 & 2.09 & 3.96\% \\
		$\operatorname{AV}(\Omega_r)~[\mathrm{rpm}]$ & 13.88 & 14.57 & 4.77\% \\
		$\operatorname{RMS}(\Omega_r)~[\mathrm{rpm}]$ & 13.94 & 14.66 & 4.93\% \\
		\bottomrule
	\end{tabular}
	\label{tab1}
\end{table}

 \subsection{Pitch controller validation}

As depicted in Fig. \ref{fig4}, the turbulent wind conditions in this specific load scenario subject the rotor to a critical state, concurrently leading to heightened fluctuations in the rotor speed curve. Consequently, a pitch controller  has been designed in \eqref{eq-40} for the regulation of rotor angular speed $\Omega_r$ in Region III.
The wind-wave conditions in this test is as same as in Fig. \ref{fig3}.
The controller parameters are specified  as $k=4.5$, $c=2$, $k_c=10$, $N=50$, $l_w=1e-3$, and $k_w=1e-3$. Additionally, the efficacy of the proposed control approach is evaluated against the baseline gain-scheduled proportional-integral (GSPI) controller \cite{jonkman2009definition} and the adaptive super-twisting (ASTW) controller introduced in \cite{zhang2021individual}.
\begin{figure}[!htbp]
	\centering
		\includegraphics[width=0.47\textwidth]{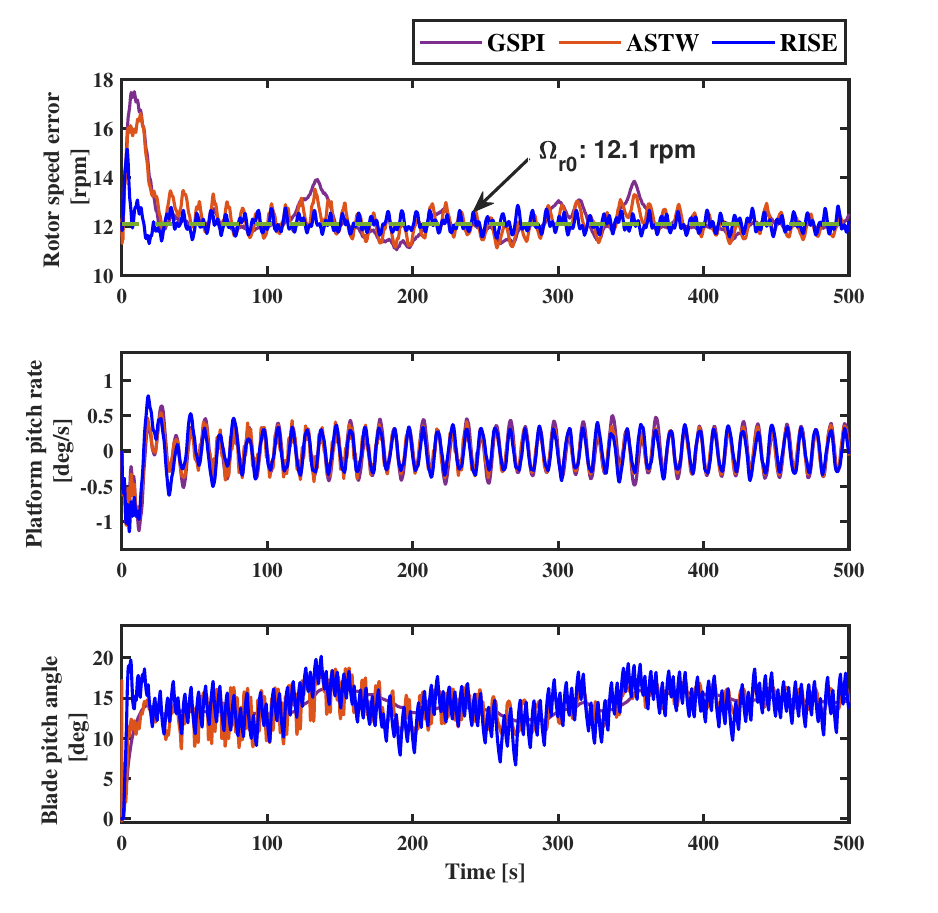}
	\caption{Dynamic responses of rotor speed, platform pitch rate and blade pitch angle.}
	\label{fig5}
\end{figure}

\begin{figure}[!htbp]
	\centering
		\includegraphics[width=0.45\textwidth]{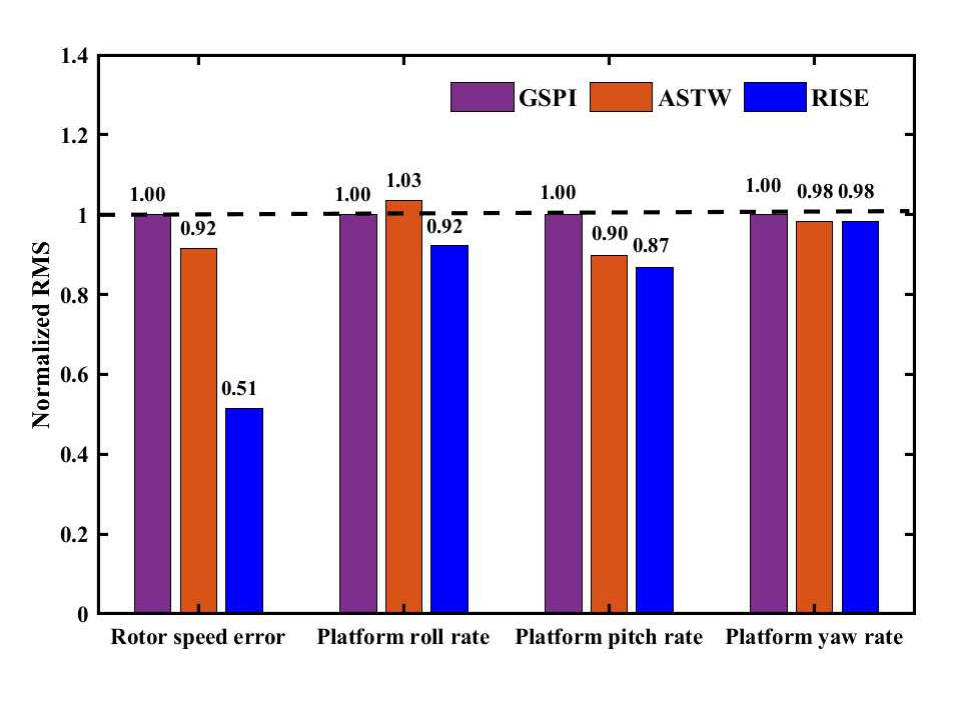}
	\caption{The normalized RMS values.}
	\label{fig6}
\end{figure}

Fig. \ref{fig5} displays the comparison results for rotor speed, platform pitch rate, and blade pitch angle.
To comprehensively assess the regulatory and loads mitigation performance, this study employs normalized RMS values for rotor speed error and platform motion rates along roll, pitch, and yaw axes, as depicted in Fig. \ref{fig6}. Simultaneously, Table \ref{tab2} presents the damage equivalent loads (DEL) at the tower base (TB), blade root (BR), fair-lead force (FF), and anchor force (AF) for the three mooring lines.
 The outcomes in Figs. \ref{fig5}--\ref{fig6} reveal that our control scheme effectively maintains rotor speed around its rated value with minimal fluctuations.
 Specifically, compared with GSPI and ASTW   methods, the RMS values for rotor speed error of the proposed method is decreased by 49.00\% and 44.56\%, respectively.
 Further, the platform roll and pitch rates are reduced by 11.96\% and 3.45\%, respectively, compared to the ASTW method.
 From Table \ref{tab2}, it is noteworthy that the ASTW and RISE controllers exhibit higher DEL of BR than the GSPI method, indicating a more pronounced demand on the blade pitch actuator.
 While the suggested approach does not enhance DEL of BR moments  in comparison to the GSPI method, it notably enhances the DELs for tower base moments and two mooring lines. This improvement is also crucial for ensuring the stability and viability of the SSFWT system.

\begin{table}[!htbp]
	\centering
\caption{ Normalized  DEL values.}
	\renewcommand{\arraystretch}{1.2} %
	\begin{tabular}{
			c
			S[table-format=2.4]
			S[table-format=2.4]
			S[table-format=2.4]
		}
		\toprule\toprule
		& \textbf{GSPI} & \textbf{ASTW} & \textbf{RISE} \\
		\midrule
		  \textbf{TB} & 1.00 & 1.04 & 0.95\\
        \textbf{BR} &  1.00 & 1.00& 1.00 \\
        $\textbf{FF}_1$& 1.00 & 1.03 & 0.94 \\
        $\textbf{AF}_1$& 1.00 & 1.05  & 0.95 \\
        $\textbf{FF}_2$ & 1.00 & 1.05 &0.99\\
        $\textbf{AF}_2$ & 1.00 & 1.08 & 0.98 \\
        $\textbf{FF}_3$& 1.00 & 1.08 & 1.03 \\
       $\textbf{AF}_3$ & 1.00 & 1.09 & 1.04 \\
		\bottomrule
	\end{tabular}
	\label{tab2}
\end{table}

\section{Conclusion}
In this paper, a   7-DOF  reduced-order model for  SSFWT has been developed, incorporating the coupled dynamics of its external key components.
Different from the exist modeling methods, both the rotor dynamics and finite rotations of floating platform are considered in our method.
Further, based on the established model,
a novel robust nonlinear controller is proposed specifically designed to achieve power production regulation while  mitigating platform pitch motion for SSFWT under wind and wave disturbances.
Moreover, an OLA is utilized to capture the unknown dynamics resulting from unmeasurable external environmental inputs and to address model uncertainties arising from time-varying parameters.
Finally, the accuracy of the proposed model  against FAST model is validated, while the superiority of
the proposed pitch controller are clearly demonstrated through comprehensive comparative studies with traditional pitch controllers.

\appendix
\subsection{Added mass matrix.}\label{app1}
\begin{equation}
	\mathbf{M}_{\mathrm{a}}=\mathbf{M}_{\mathrm{a1}}+\mathbf{M}_{\mathrm{a2}}
\end{equation}
with
\begin{equation*}
	\begin{aligned}
		\mathbf{M}_{\mathrm{a1}}=\sum_{i=1}^{7}\begin{bmatrix}
			M_{i_1}&-M_{i_2}\\
			\bar{M}_{i_3}&\bar{M}_{i_4}
		\end{bmatrix},\quad \mathbf{M}_{\mathrm{a2}}=\sum_{j=5}^{7}\begin{bmatrix}
			N_{j_1}&N_{j_2}\\
			N_{j_3}&N_{j_4}
		\end{bmatrix}
	\end{aligned}
\end{equation*}
where $M_{i_1}=C_{Ai}L_iT_n, ~ M_{i_2}=	C_{Ai}T_n\tilde{\mathbf{r}}_{\mathrm{bi}}L_i+M_{i3}$,
\begin{equation*}
	\begin{aligned}
		 & M_{i_3}=C_{Ai}\int^{z_t}_{z_b}\begin{bmatrix}
			0&-z&0\\
			z&0&0\\
			0&0&0
		\end{bmatrix}~dz,~\\
&\bar{M}_{i_3}=\tilde{\mathbf{r}}_{\mathrm{bi}}M_{i1}+M_{i3},\\
   &	\bar{M}_{i_4}=-\tilde{\mathbf{r}}_{\mathrm{bi}}M_2-M_4+M_5,~
	M_{i_4}= M_{i_3} \tilde{\mathbf{r}}_{\mathrm{bi}},\\	
		&M_{i_5}=C_{Ai}\int^{z_t}_{z_b}\begin{bmatrix}
			z^2&0&0\\
			0&z^2&0\\
			0&0&0
		\end{bmatrix}~dz
	\end{aligned}
\end{equation*}
and
\begin{equation*}
	\begin{aligned}
		&N_{j_1}=C_{A_zj}T_h, \quad N_{j_2}=-C_{A_zj}T_h\tilde{\mathbf{r}}_{\mathrm{c_j}},\\
		&N_{j_3}=\tilde{\mathbf{r}}_{\mathrm{c_j}}N_{j_1}, \quad N_{j_4}=\tilde{\mathbf{r}}_{\mathrm{c_j}}N_{j_2}
	\end{aligned}
\end{equation*}
with
\[
T_n=\begin{bmatrix}
	1&0&0\\
	0&1&0\\
	0&0&0
\end{bmatrix},\quad
T_h=\begin{bmatrix}
	0&0&0\\
	0&0&0\\
	0&0&1
\end{bmatrix}
\]
 In order to simplify the expression of added mass, we assume that it does not get recomputed as the lengths of the submerged cylinders change, i.e., $L_i$ equals to its original value. Then we have
 	\begin{equation*}
	\mathbf{M}_{a}=  \begin{bmatrix}
		   b_{11} & 0 & 0 & 0 & b_{15} & 0 \\
		      & b_{11} & 0 & -b_{15} & 0 &0 \\
		     &  & b_{33} & 0 & 0 & 0 \\
		      &   &    & b_{44} & 0 & 0 \\
		     & * &  &  & b_{55} & 0 \\
		      &   &  &  &  & b_{66}
		\end{bmatrix}
		\end{equation*}
		where $b_{11}=\sum_{i=1}^{7} C_{Ai}L_i, ~b_{33}=\sum_{j=5}^{7}C_{A_zj}, ~b_{15}= \sum_{i=1}^{7} C_{A i}L_i^2(1 + 2 r_{bi3})/2 $, $b_{66}=C_{Ai}L_ir_{b_{i1}}^2 + C_{Ai}L_ir_{b_{i2}}^2$,
  $b_{44}=b_{55}=  \sum_{i=1}^{7}\sum_{j=5}^{7}(  C_{A i}L_i^3/3   + C_{A i}L_i^2r_{bi3} + C_{A i}L_i r_{bi3}^2 + C_{A_z j}r_{cj 1}^2)$.

 \subsection{System matrices in \eqref{eq-23}.} \label{app2}
\begin{equation*}
		\begin{aligned}
\mathbf{\bar{M}} =\begin{bmatrix}
				\mathbf{I}_{6\times6}& \mathbf{0}_{6\times7}\\
				\mathbf{0}_{7\times6}& \mathbf{\bar{M}}_s
			\end{bmatrix}
		\end{aligned}
	\end{equation*}
	where $\mathbf{\bar{M}}_s  = \begin{bmatrix}
				\mathbf{M}_{s1}&\mathbf{M}_{s2}\\
				\mathbf{M}_{s3}&\mathbf{M}_{s4}
			\end{bmatrix} +\begin{bmatrix}
				\mathbf{M}_a &\mathbf{0}_{6\times1}\\
				\mathbf{0}_{1\times6}&0
			\end{bmatrix}$,

	\begin{equation*}
	\begin{aligned}
		& \mathbf{M}_{s1}= \begin{bmatrix}
			a_{1} & 0 & 0 & 0 & a_{2} & 0   \\
			& a_{1} & 0 & -a_{2} & 0 & a_{3}   \\
			& &a_{1} & 0 & -a_{3} & 0  \\
			& & & a_{4} & 0 &a_{6}   \\
			& *& & &a_{5} & 0  \\
			& & & & &a_{7}
		\end{bmatrix}\\
		& 	\mathbf{M}_{s2}^T=\mathbf{M}_{s3}=\begin{bmatrix}
			0 & 0 & 0 & I_{r x} & 0 & 0
		\end{bmatrix},~ \mathbf{M}_{s4}=I_{rx}
	\end{aligned}
\end{equation*}
	with
	\begin{equation*}
		\begin{aligned}
			a_{1} & =m_p+m_t+m_{n c}+m_r,\\
			a_{2}&=H_r(m_r+m_{n c})+m_t H_{t}, \\
			a_{3} & =h_{nc} m_{n c}-h_r m_r,\\
			a_{4}&=I_{p x}+I_{t x}+I_{n c x}+I_{r x}+H_r^2 (m_r+m_{n c})+H_{t}^2 m_t, \\
			a_{5} & =I_{p y}+I_{t y}+I_{n c y}+I_{r y}+m_r (H_r^2+h_r^2 )\\
                      &\quad+m_{n c}(H_r^2+h_{nc}^2 )+H_{t}^2 m_t, \\
			a_{6} & =H_r h_r m_r-H_r h_{nc} m_{n c},\\
			a_{7} &=I_{p z}+I_{t z}+I_{n c z}+I_{r z}+m_r h_r^2+m_{n c} h_{nc}^2
		\end{aligned}
	\end{equation*}
	and
	\begin{equation*}
		\begin{aligned}
	&\mathbf{s}(\mathbf{r}_\mathrm{p}, \boldsymbol{\theta}_\mathrm{p}, \mathbf{v}_\mathrm{p}, \boldsymbol{\omega}_\mathrm{p}, \Omega_r)\\
&=-\begin{bmatrix}
				\mathbf{0}_{6\times6}& \mathbf{0}_{6\times7}\\
				\mathbf{0}_{7\times6}& \mathbf{C}_s
			\end{bmatrix} \begin{bmatrix}
	[\mathbf{r}_\mathrm{p}]_I\\
	[\boldsymbol{\theta}_\mathrm{p}]_I\\
	[\mathbf{v}_\mathrm{p}]_p\\
	[\boldsymbol{\omega}_\mathrm{p}]_p\\
	\Omega_r
\end{bmatrix}+ g\begin{bmatrix}
							a_{1} s_{\theta_y} \\
						-a_{1} c_{\theta_y} s_{\theta_x} \\
						-a_{1} c_{\theta_x} c_{\theta_y}\\
						c_{\theta_y} s_{\theta_x}	a_{2} \\
						s_{\theta_y}	a_{2}+c_\phi c_{\theta_y}m_d \\
						-s_{\theta_x} c_{\theta_y}m_d\\
					\end{bmatrix}
			\end{aligned}
	\end{equation*}
where $m_d=h_{nc} m_{n c}+h_r m_r$, and $\mathbf{C}_{\mathrm{s}}=	\begin{bmatrix}
				\mathbf{C}_{s1}&\mathbf{C}_{s2}\\
				\mathbf{C}_{s3}&\mathbf{C}_{s4}
			\end{bmatrix}$, with
	\begin{equation*}
		\begin{aligned}
			&\mathbf{C}_{s1} =\widetilde{S} \mathbf{M}_{s1} =
			\begin{bmatrix}
				[\widetilde{\boldsymbol{\omega}}_\mathrm{p}]_p & \mathbf{0}_{3 \times 3}   \\
				[\widetilde{\boldsymbol{v}}_\mathrm{p}]_p & [\widetilde{\boldsymbol{\omega}}_\mathrm{p}]_p
			\end{bmatrix}\mathbf{M}_{s1}\\
			&\mathbf{C}_{s2} =\begin{bmatrix}
				0 & 0 & 0 & 0 & I_{r x} \omega_z & -I_{r x} \omega_y
			\end{bmatrix}^\mathrm{T}\\
			&\mathbf{C}_{s3} =\begin{bmatrix}
				0 & 0 & 0 & 0 & 0 & 0
			\end{bmatrix},~\mathbf{C}_{s4} =0
		\end{aligned}
	\end{equation*}
 
\begin{equation*}
	\begin{aligned}
		b_1=\frac{1}{I_{rx}},~~b_2=\frac{h_1(h_2 - H_r h_3)}{(h_1h_2^2 + h_3h_4^2 - h_5 h_3h_1)}
	\end{aligned}
\end{equation*}
with  $h_1=a_1+b_{33},~
h_2=a_2+ b_{15},~h_3=a_1+b_{11},~
h_4=a_3,~
h_5=a_5+b_{55}$.
Since $H_r>H_t$, it can be easily deduced that $h_2 - H_r h_3<0$. Also, by some simple  calculations, it can be  obtained that $ h_1h_2^2 + h_3h_4^2 - h_5 h_3h_1 <0$. Thus, we have $b_2>0$.

\subsection{Proof of Theorem 1.} \label{app3}

Consider the following Lyapunov function candidate:
\begin{equation}\label{eq-49}
	V_L = \frac{1}{2} \xi^T \xi+\frac{1}{2} \bar{\xi}^T T(Y) \bar{\xi}+Q+P+\frac{1}{2} \tilde{\gamma}^2
\end{equation}
with $Y=[\mathbf{X}^T,\beta,\mathbf{v}_w]^T$, and
\begin{equation}\label{eq-50}
	\begin{aligned}
		P  =H_0-\int_0^t H(\varsigma ) d \varsigma,\quad
		Q  = \frac{1}{2 l_\omega} \operatorname{tr} (\tilde{W}^T \tilde{W} )
	\end{aligned}
\end{equation}
Then, there exist two  $\mathcal{K}$-class functions $V_{k_1}$,~$V_{k_2}$, such that
\begin{equation}\label{eq-51}
V_{k_1}(M) \leq V_L \leq V_{k_2}(M)
\end{equation}
with $ M=\left[\begin{array}{lllll} \xi^T & \bar{\xi}^T & \sqrt{Q} & \sqrt{P} & \tilde{\gamma}\end{array}\right]^T $, and
\begin{equation}\label{eq-52}
	\begin{aligned}
		V_{k_1}&=\alpha_1\|M\|^2,~\alpha_1=\frac{1}{2} \min \left\{1, T(Y)\right\}>0,\\
		V_{k_2} &= \alpha_2\|M\|^2,~\alpha_2=\max \left\{1, T(Y)\right\}>0
	\end{aligned}
\end{equation}
Next, the time derivative of $V_L$ is obtained as
\begin{eqnarray*}
	\begin{aligned}
		\dot{V_L}= & \xi^T \dot{\xi}+\bar{\xi}^T T(Y) \dot{\bar{\xi}}+\frac{1}{2} \bar{\xi}^T \dot{T}(Y) \bar{\xi}+\frac{1}{l_\omega} \operatorname{tr}(\tilde{W}^T \dot{\tilde{W}}) \\
		& -\bar{\xi}^T\left(\bar{D}-\gamma_d \operatorname{sgn}(\xi)\right)-\frac{\left\|w_m\right\|^2 }{4} \xi^T \operatorname{sgn}(\xi)\\
		&+\dot{\xi}^T \bar{N}+\tilde{\gamma} \dot{\tilde{\gamma}} \\
		= & \xi^T \dot{\xi}+\bar{\xi}^T \times[-\xi-\tilde{W}^T \phi(Z)-\left(k_c+1\right) c \bar{\xi}\\
		& -\gamma \operatorname{sgn}(\xi)+\bar{D}] +c \xi^T \tilde{W}^T \phi(Z)\\
		& -k_w|\xi| \operatorname{tr}(\tilde{W}^T \hat{W}) -\bar{\xi}^T(\bar{D}\left.-\gamma_d \operatorname{sgn}(\xi)\right)\\
		&-\frac{\left\|w_m\right\|^2 }{4} \xi^T \operatorname{sgn}(\xi)+\dot{\xi}^T \bar{N}+\tilde{\gamma}^T \bar{\xi}^T(t) \operatorname{sgn}(\xi)\\
		= & \xi^T \dot{\xi}+\bar{\xi}^T\left(-\xi-\left(k_c+1\right) c \bar{\xi}\right)-\bar{\xi}^T \tilde{W}^T \phi(Z) \\
		& +c \xi^T \tilde{W}^T \phi(Z)+\dot{\xi}^T \bar{N}- k_w|\xi| \operatorname{tr}(\tilde{W}^T \tilde{W}) \\
		& - k_w|\xi| \operatorname{tr}(\tilde{W}^T W)-\frac{\left\|w_m\right\|^2 }{4} \xi^T \operatorname{sgn}(\xi) \\
		& -\gamma \bar{\xi}^T \operatorname{sgn}(\xi)+\gamma_d \bar{\xi}^T \operatorname{sgn}(\xi)+\tilde{\gamma} \tilde{\xi}^T \operatorname{sgn}(\xi) \\
		=& \xi^T \dot{\xi}+\bar{\xi}^T\left(-\xi-\left(k_c+1\right) c \bar{\xi}\right)- k_w|\xi| \operatorname{tr}(\tilde{W}^T \tilde{W}) \\
		& - k_w|\xi| \operatorname{tr}(\tilde{W}^T W)-\frac{\left\|w_m\right\|^2 }{4}|\xi|
	\end{aligned}
\end{eqnarray*}
Further, it derives
\begin{eqnarray*}
	\begin{aligned}
		V_L\leq & \xi^T \dot{\xi}+\bar{\xi}^T (-\xi- (k_c+1) c
		\bar{\xi} )- k_w|\xi| (\|\tilde{W}\| \\
		&- \|W\|/2 )^2+\frac{ k_w|\xi|\|W\|^2}{4}-\frac{\left\|w_m\right\|^2 }{4}|\xi|  \\
		\leq & \xi^T \dot{\xi}+\bar{\xi}^T (-\xi- (k_c+1 ) c \bar{\xi} )- k_w|\xi| (\|\tilde{W}\|\\
		 &- \|W\|/2 )^2\\
\leq&-c \xi^T \xi- (k_c+1 )c \bar{\xi}^{T}\bar{\xi}
	\end{aligned}
\end{eqnarray*}
which leads to
\begin{eqnarray}
	\dot{V_L} \leq-L(M)
\end{eqnarray}
with $L(M)=c\left\|\left[\begin{array}{rr}\xi^T & \bar{\xi}^T\end{array}\right]\right\|^2$, which is a continuously positive semi-definite function defined  over an arbitrary compact set $\Gamma_M \subset \mathbb{R}^{5}$.
Based on \eqref{eq-51} and \eqref{eq-52}, it can be inferred that $V_L$ is bounded over $\Gamma_M$. Therefore, $\xi$, $\bar{\xi}$, $\tilde{W}$, and $\tilde{\gamma}$ are all bounded. By recalling the definition of $\xi$ in \eqref{eq-29}, it is evident that both $\Delta\Omega_{r}$ and $\Delta \dot{\theta}_y$ are bounded.
In addition, the boundedness of $\bar{\xi}$ implies that $\dot{\xi}$ is also bounded.
Note that the unknown disturbance $d(t)$ and its   time derivatives are generally assumed to be bounded. Then from  \eqref{eq-30}, it derives that the control input $\beta$ is bounded.

Utilizing conventional linear analysis techniques, it can be inferred that the variables in equations \eqref{eq-33}, \eqref{eq-36}, \eqref{eq-39}, and \eqref{eq-42} exhibit bounded behavior. Consequently, all signals within the closed-loop system are ensured to remain within boundsHence, it can be  deduced that $L(M)$ is uniformly continuous.
Moreover, Barbalat's lemma \cite{c10} can be applied to deduce that
\begin{eqnarray}
	|\xi(t)|^2 \rightarrow 0 \quad \text{and} \quad |\bar{\xi}(t)|^2 \rightarrow 0 \quad \text{as} \quad t \rightarrow \infty.
\end{eqnarray}
This can be further obtained that $ |\xi(t) | \rightarrow 0$ as $t \rightarrow \infty$.

\bibliographystyle{IEEEtran}
\bibliography{refs}

\end{document}